\documentclass[11pt]{article}
\usepackage{amssymb, amsmath, amsthm}
\usepackage{a4wide}
\usepackage{xcolor}
\usepackage{prettyref}
\usepackage{algorithm}

\usepackage[colorlinks=true]{hyperref}
\usepackage[hyperpageref]{backref}
\usepackage[capitalize,nameinlink,noabbrev]{cleveref}
\hypersetup{
	linkcolor={teal!100!black},
	citecolor={teal!100!black},
	urlcolor={teal!100!black},
    pdftitle={Lower Bounds for Adaptive Relaxed Locally Decodable Codes via the Capped Horvitz--Thompson Estimator},
}

\newtheorem{theorem}{Theorem}[section]
\newtheorem{lemma}[theorem]{Lemma}
\newtheorem{construction}[theorem]{Construction}
\newtheorem{claim}[theorem]{Claim}
\newtheorem{fact}[theorem]{Fact}
\newtheorem{corollary}[theorem]{Corollary}
\newtheorem{definition}[theorem]{Definition}
\newtheorem{remark}[theorem]{Remark}
\newtheorem{observation}[theorem]{Observation}
\newtheorem{thm}{Theorem}

\def\R{\mathbb{R}}
\def\D{\mathcal{D}}
\def\P{\mathcal{P}}
\def\B{\mathcal{B}}

\newcommand{\supp}{\mathrm{supp}}
\newcommand{\dist}{\mathrm{dist}}

\DeclareMathOperator{\E}{\mathbb{E}}
\DeclareMathOperator{\Var}{Var}

\newcommand{\savehyperref}[2]{\texorpdfstring{\hyperref[#1]{#2}}{#2}}

\newrefformat{section}{\savehyperref{#1}{Section~\ref*{#1}}}
\newrefformat{lemma}{\savehyperref{#1}{Lemma~\ref*{#1}}}
\newrefformat{fact}{\savehyperref{#1}{Fact~\ref*{#1}}}
\newrefformat{claim}{\savehyperref{#1}{Claim~\ref*{#1}}}
\newrefformat{remark}{\savehyperref{#1}{Remark~\ref*{#1}}}
\newrefformat{construction}{\savehyperref{#1}{Construction~\ref*{#1}}}
\newrefformat{corollary}{\savehyperref{#1}{Corollary~\ref*{#1}}}
\newrefformat{thm}{\savehyperref{#1}{Theorem~\ref*{#1}}}
\newrefformat{algo}{\savehyperref{#1}{Algorithm~\ref*{#1}}}
\newcommand{\pref}{\prettyref}
\newrefformat{equation}{Equation~\ref{#1}}

\title{Lower Bounds for Adaptive Relaxed Locally Decodable Codes via the Capped Horvitz--Thompson Estimator}
\author{Guy Goldberg \thanks{GG is supported by ERC Consolidator Grant 772839 and ISF Grant 2073/21.} \\Weizmann Institute of Science \\ \textsf{guy.goldberg@weizmann.ac.il} \and
Tom Gur\thanks{TG is supported by ERC Starting Grant 101163189 and UKRI Future Leaders Fellowship MR/X023583/1.}\\University of Cambridge\\ \textsf{tg508@cam.ac.uk} \and
Sidhant Saraogi\thanks{SS is supported by the NSF CAREER grant CCF-1845125.} \\Georgetown University \\ \textsf{ss4456@georgetown.edu}}
\date{November 5, 2026}

\begin{document}
\maketitle
\begin{abstract}
    We prove a nearly optimal lower bound on the blocklength of general relaxed locally decodable codes (RLDCs), extending the result of~\cite{GGS26} to nonlinear encodings and adaptive decoders with two-sided error.
    Namely, we show that any $q$-query RLDC $C\colon \{0,1\}^k \to \{0,1\}^n$ must satisfy $n = k^{1+\Omega(1/q)}$, closely matching the upper bound of $n = k^{1+O(1/q)}$ by Ben-Sasson, Goldreich, Harsha, Sudan, and Vadhan (STOC 2004).
    
   Our proof reduces showing RLDC lower bounds to estimating the output distribution of adaptive decoders.
   To this end, we introduce a capped Horvitz--Thompson estimator with uniformly bounded discrete derivatives on the $p$-biased cube. We establish its concentration of measure via self-bounding inequalities and control the downward bias induced by capping through a spreadness argument.
\end{abstract}
\thispagestyle{empty}
\newpage

\setcounter{page}{1}

⊥

\section{Introduction}
    The notion of relaxed locally decodable codes (RLDCs) is a well-studied relaxation of locally decodable codes, which allows the decoder to abort on corrupted inputs while still requiring it to correctly decode valid codewords.
    
    More precisely, a $q$-RLDC is an error-correcting code $C: \{0, 1\}^k \rightarrow \{0, 1\}^n$ that admits a randomized decoder $D$ that recovers individual message bits using at most $q$ queries to a purported codeword $w \in \{0, 1\}^n$.
    On a valid codeword $w=C(x)$ and index $i$, the decoder returns $x_i$ with probability at least $2/3$.
    On a noisy codeword within the decoding radius, it must either return $x_i$ or an abort symbol $\bot$ with probability at least $2/3$.

    In a seminal work, Ben-Sasson, Goldreich, Harsha, Sudan, and Vadhan~\cite{BGHSV06} introduced this relaxation and obtained $q$-RLDCs with blocklength $n = k^{1+O(1/q)}$ (see also \cite{Goldreich24}).
    This implies $O(1)$-RLDCs with nearly linear length, whereas the best known $O(1)$-query locally decodable codes have superpolynomial length \cite{yekhanin08,efr12}.
    Despite receiving much attention~\cite{GR18, GG18, GGK19, GRR20,AS21, GG21, DGMT22, CGS22, CY22, Goldreich24, KM24, CY24,GMWZ26}, this original construction still achieves the best parameters for $O(1)$-RLDCs.

    This lack of progress is partially explained by recent lower bounds.
    For constant-query RLDCs, Dall’Agnol, Gur, and Lachish \cite{DGL23} (building on \cite{gl19}) have shown the general lower bound $n=k^{1+\widetilde\Omega(1/q^{2})}$.
    More recently, a nearly optimal lower bound of $n=k^{1+\Omega(1/q)}$ was shown for RLDCs with \emph{non-adaptive decoders} and perfect completeness \cite{GGS26}.
    Using the reduction of \cite{Goldberg24}, this bound also extends to all linear RLDCs.

    Nevertheless, the question of the optimal lower bound for general RLDCs remained open.
    In particular, it was unknown whether adaptivity and non-linearity allow $q$-query RLDCs of length $n=k^{1+o(1/q)}$.
    
    In this work, we resolve this question by proving a nearly optimal lower bound on the blocklength of \emph{any} RLDC.
    
    \begin{thm} \label{thm:main-intro}
        Suppose $C :\{0,1\}^k\rightarrow \{0, 1\}^n$ is a $q$-RLDC with constant decoding radius.
        Then, for every constant $\varepsilon > 0$,
        \[
            n=\Omega_{\varepsilon} \left(q^{-2} \left(\frac{k}{\log k}\right)^{1+\frac{1}{(2 + \varepsilon)q}} \right) \; .
        \]
    \end{thm}
    This recovers the nearly optimal lower bound in~\cite{GGS26}, without assuming either non-adaptivity or perfect completeness.
    In particular, in the constant-query regime, \cref{thm:main-intro} gives a bound of $n=k^{1+\Omega(1/q)}$, matching the exponent of the upper bound in \cite{BGHSV06} up to constant factors.
    In the constant-rate regime, this implies a lower bound of $q=\Omega(\log k/\log\log k)$.
    The result extends to completeness and soundness probabilities other than $2/3$.
    See \pref{thm:main} for the precise statement.

\subsection{The Capped Horvitz--Thompson Estimator} \label{section:intro-cht}

    The proof of \pref{thm:main-intro} introduces the notion of the \emph{capped Horvitz--Thompson Estimator}, which may be of independent interest.
    As we discuss in \cref{sec:tech}, our approach to proving the adaptive RLDC lower bound relies on estimating the total mass of certain sub-distributions over query sets, using only binomially sampled input coordinates.
    
    More precisely, consider a sub-distribution $\mu$ over subsets of $[n]$.
    Let $W \subseteq [n]$ be obtained by sampling each coordinate independently with probability $p$.
    We would like to estimate the total mass $\sum_S\mu(S)$, including the contribution of non-sampled sets $S \not\subseteq W$, using only the weights $\mu(S)$ of sets $S \subseteq W$.
    In our application, this total mass corresponds to the probability that the decoder produces a specified output on a specified input string.
    
    The \emph{Horvitz--Thompson estimator} \cite{HT52} is a classical statistical tool used to obtain an unbiased estimate of the total mass by assigning each sampled set $S$ the weight of $\mu(S)p^{-|S|}$, compensating for its inclusion probability $p^{|S|}$. 
    Although unbiased, this estimator need not concentrate around its mean, as overlapping sets can produce highly dependent contributions, and a single sampled coordinate may have a large influence on the estimate.

    To address this, we impose a cap on the marginal contribution of each coordinate.
    The resulting $\Gamma$-capped Horvitz--Thompson (CHT) estimator bounds the discrete derivatives so that adding or removing one coordinate from the sample changes the estimate by at most $\Gamma$, while preserving as much of the original weight as these constraints permit.
    In the following definition we identify the sampled set $W$ with its indicator vector $x = \mathbf{1}_W \in \{0, 1\}^n$.
\begin{definition}
\label{def:CHTintro}
    Let $\mu$ be a sub-distribution over subsets of $[n]$, and fix $p \in (0,1]$ and $\Gamma > 0$. For $x \in \{0,1\}^n$, write $x_S = \prod_{i \in S} x_i$.
    The Horvitz--Thompson estimator is given by
    \[R_{\mu,p}(x) = \sum_{S \subseteq[n]} \mu(S) p^{-|S|} x_S.\]
    The \emph{$\Gamma$-Capped $p$-Horvitz--Thompson estimator} $\overline{R}_{\mu,p,\Gamma}(x)$ is the maximum total weight $\sum_{S \subseteq [n]} w_S$ over assignments of weights
    \[
        0 \leq w_S \leq \mu(S)p^{-|S|}x_S \; ,
    \]
    satisfying $\sum_{S \ni i} w_S \leq \Gamma$ for every $i \in [n]$.
\end{definition}

    Using concentration inequalities for self-bounding functions, we show that the capped Horvitz--Thompson estimator concentrates around its mean, with bounds depending on the capping parameter $\Gamma$.
    The tension is, however, that the capping also induces a downward bias of the estimator, which can be large for general sub-distributions.
    We show that if the underlying sub-distribution is \emph{spread}, then this bias is small.
    Together, these results show that the CHT estimator concentrates around its mean for any sub-distribution, and that this mean is close to the total mass when the sub-distribution is sufficiently spread.

    \paragraph{An analytic perspective. }
    The Horvitz--Thompson estimator can be viewed as a multilinear polynomial with nonnegative coefficients.
    The CHT estimator can be interpreted as the pointwise maximum of such polynomials whose derivatives are uniformly bounded.
    The downward bias is therefore exactly the $\ell_1$ distance between the two functions on the $p$-biased cube.
    We believe this perspective suggests interesting connections to Fourier analysis on the $p$-biased cube, which we leave for future work.
 
\subsection{Related Work}
    While the state of the art of constant-query RLDCs is still the original work of Ben-Sasson, Goldreich, Harsha, Sudan, and
    Vadhan~\cite{BGHSV06}, which was observed in \cite{Goldreich24} to achieve blocklength $n=k^{1+O(1/q)}$, much work has been done on constructions of RLDCs and closely related notions.
    This includes RLDCs that are also locally testable \cite{GR18,GGK19}, and relaxed locally correctable codes (RLCCs)~\cite{GRR20,CGS22,AS21,CY22}, where the decoder recovers individual \emph{codeword} symbols. In the constant-rate regime, Kumar and Mon~\cite{KM24} constructed RLCCs with polylogarithmic query complexity. Cohen and Yankovitz~\cite{CY24} subsequently optimized the query complexity to $(\log n)^{2+o(1)}$. These constructions are linear, and hence also yield RLDCs, leaving a gap to our lower bound of $q=\Omega(\log k/\log\log k)$.

In terms of lower bounds, our proof follows the sample-based decoding approach underlying previous RLDC lower bounds~\cite{gl19,DGL23,Goldreich23,GGS26}. Goldberg~\cite{Goldberg24} showed that for the special case of linear codes, it is possible to convert arbitrary decoders to non-adaptive ones with perfect completeness, with only minor overhead in query complexity. Consequently, the nearly tight lower bound of~\cite{GGS26} applies both to non-adaptive RLDCs with perfect completeness and to general linear RLDCs.

\subsection*{AI Disclosure}
The main conceptual and technical ideas in this paper are due to the authors. 
We used ChatGPT (GPT-6 Astra and GPT-5.6 Sol) in proofreading, literature review, minor stylistic editing, and assistance with certain technical claims. More specifically, we used it in the development of \pref{lemma:cht-expectation}, where it helped us derive a first-principles bound on the expectation gap, instead of relying on the Kim--Vu inequality, which would incur an additional $\log^{2q} k$ loss in the resulting lower bound. We also used it in deriving the proof of the technical lemma \pref{lemma:var}, and in working out details of the counterexample in \pref{section:adaptive-challange}.
The authors verified the correctness and originality of all content in this work.

\section{Proof Overview}
\label{sec:tech}
    Our starting point is the \emph{sample-based decoder} paradigm for showing RLDC lower bounds.
    We first review this paradigm and the obstacles that arise for adaptive decoders with two-sided error, and then describe how our approach overcomes them using the Capped Horvitz--Thompson estimator.

\subsection{From Sample-Based Decoders to Lower Bounds}
    The core of the lower bound proof is an information-theoretic argument.
    By Fano's inequality, recovering $k$ independent bits, each with error probability at most $\varepsilon < 1/2$ requires observing $(1-h_2(\varepsilon))k =\Omega(k)$ bits of information. 
    
    In this work, we introduce a new notion of a \emph{sample-based decoder}.
    Such a decoder gets sample access to a valid codeword $C(x)$, and it must recover each bit of $x$ with error probability at most $\varepsilon < 1/2$.
    Since the sample-based decoder observes, in expectation, $pn$ codeword symbols, the argument above implies that if a code $C$ has a sample-based decoder with sampling probability $p$, then $pn = \Omega(k)$.

    More precisely, let $X$ be uniformly distributed over $\{0, 1\}^k$, and let $W \subseteq [n]$ be the random set of sampled coordinates.
    Intuitively, recovering each bit with error probability at most $\varepsilon$ means that after the sample-based decoder observes $(W, C(X)|_W)$, at most $kh_2(\varepsilon)$ bits of uncertainty remain about $X$.
    More precisely, we use Fano's inequality and the subadditivity of conditional entropy to prove that $H(X \mid W, C(X)|_W) \leq k h_2(\varepsilon)$.     On the other hand, since $H(X) = k$, observing an expected number of $pn$ codeword bits leaves at least $k-pn$ bits of uncertainty about $X$.
    Namely, $H(X \mid W, C(X)|_W) \geq k-np$.
    Combining the bounds yields the result.

    This global-decoding paradigm has been used in all previous RLDC lower bounds.
    However, previous work imposed the stronger requirement that the decoder must recover all of the bits of $x$ \emph{simultaneously} with high probability. 
    We view the introduction of this weaker notion of a sample-based decoder as a conceptual contribution, as it is the natural information-theoretic formulation of sample-based decoding.
    
    Besides isolating the minimal guarantee needed for the lower-bound argument, it allows us to further improve the parameters of the lower bound of \cite{GGS26} and simplify the logic of the sample-based decoder.
    Specifically, ensuring simultaneous recovery of all $k$ message bits via a union bound requires reducing the failure probability for each bit to $O(1/k)$, incurring an additional logarithmic factor in the sampling probability.
    In contrast, our coordinate-wise guarantee avoids union bound and the associated logarithmic factor.
    This improves the base of the exponent in the lower bound from $\frac{k}{\log^2 k}$ in \cite{GGS26} to $\frac{k}{\log k}$ in this work.

\subsection{The Non-Adaptive, 1-Sided Error Case}
    We begin by recalling the sample-based decoding strategy of \cite{gl19, GGS26}.
    
    Consider a relaxed decoder for a code $C:\{0, 1\}^k \to \{0, 1\}^n$.
    If it is non-adaptive, and never errs on valid codewords, then without loss of generality it can be fully described by a set of distributions $\{\mu_i\}_{i \in [k]}$, each over subsets of $[n]$.
    To decode $x_i$ from a corrupted codeword $y \in \{0, 1\}^n$, the relaxed decoder samples a set $S$ according to $\mu_i$ and queries the indices in $S$.
    It then outputs according to the local view $y|_S$: if there exists $x \in \{0, 1\}^k$ consistent with the local view (that is, $C(x)|_S = y|_S$), it outputs $x_i$, and otherwise it outputs $\bot$.\footnote{It is impossible that there exists $x,x'$ with $C(x)|_S = C(x')|_S =y|_S$ and $x_i \neq x'_i$, since this would contradict the 1-sided error of the decoder.
    If there is no codeword consistent with $y|_S$, then WLOG the decoder outputs $\bot$, since any other symbol can only increase its soundness error.}

    If the sample-based decoder samples all the indices of some set $S$ from the support of $\mu_i$, it can recover $x_i$ with certainty by invoking the relaxed decoder's logic on $S$.
    Since the input is a valid codeword, the relaxed decoder is guaranteed to decode correctly.
    However, the decoder's queries may be non-smooth.
    For instance, to decode index $i$, the decoder might always query index $i$ itself.
    In this case, the sample-based decoder must sample index $i$ to capture any complete query-set.
    Since this happens only with probability $p$, obtaining a constant success probability would require $p=\Omega(1)$, yielding only a trivial linear lower bound.

    \paragraph{The sample-based decoder. }
    To solve this, \cite{gl19, GGS26} use the following idea:
    to decode $x_i$, they first identify the heavy indices, i.e., indices that the total mass of sets that include them (according to $\mu = \mu_i$) is above some threshold $m$.
    Let us use $\langle u \rangle = \{S \subseteq [n] \mid u \in S\}$ to denote all the sets that contain $u$, and $H = H_m(\mu) := \{u \in [n] \mid \mu(\langle u \rangle) \geq m \}$ is the set of heavy indices.
    Then, instead of trying to sample the indices in $H$, \emph{guess} the values of $C(x)$ at $H$.
    By a standard counting argument, and since each set in $\supp(\mu_i)$ is of size at most $q$, we have $|H| \leq \frac{q}{m}$.

    Specifically, the sample-based decoder of \cite{gl19} and \cite{GGS26} works as follows, for a parameter $m > 0$.
    First, it samples each bit of its input codeword with probability $p$, to get a set of sampled indices  $W \subseteq [n]$ and a partial view of the codeword $c|_W$.
    Next, to decode each $x_i$, it identifies $H_m$ as discussed above.
    For each assignment $\kappa \in \{0, 1\}^H$ (a ``guess''), it checks all of the sets $S \in \supp(\mu)$ such that $S \subseteq W \cup H$.
    Those are the ``local views'' that the decoder can simulate.
    It then feeds all of those local views into the relaxed decoder, using $c$ for the sampled indices $W$, and $\kappa$ for the indices in $H$.

    If there is a consensus (i.e., at least one local view is simulated, and all local views lead the decoder to output the same bit), it outputs that bit as the value of $x_i$.

    \paragraph{Analysis. }
    
    In the correct iteration, where $\kappa$ is consistent with $C(x)$, all simulated local views are also consistent with the correct codeword, and therefore, by perfect completeness, yield the correct output bit.

    In every other iteration, the simulated views are consistent with a corrupted codeword differing from $C(x)$ on at most $|H|$ coordinates.
    If $|H| \leq \delta n$, soundness guarantees that a large fraction (with respect to $\mu$) of the query sets are ``good'', meaning that they lead the decoder to output the correct bit or $\bot$.

    The sample-based decoder can output an incorrect bit only if, in some iteration, none of the simulated views corresponds to a ``good'' query set.
    Thus, denoting the good query sets as $G \subseteq \supp(\mu)$, we seek a threshold $m$ for which
    \begin{equation} \label{equation:soundness}
        \Pr[\exists S \in G \text{ such that } S \subseteq W \cup H] \geq 1-\exp(- \Omega(|H|)) \; ,
    \end{equation}
    where the probability is over the choice of $W$.
    This failure bound allows us to take a union bound over all $2^{|H|}$ assignments to $H$, showing that, with constant probability, the sample-based decoder does not output an incorrect bit in any iteration.
    
    \paragraph{Finding a suitable threshold. }    
    There is a tradeoff in choosing the threshold $m$, which determines the size of $H$:
    a larger $H$ makes the requirement $S \subseteq W \cup H$ easier to satisfy, but also requires a smaller failure probability.

    While \cite{gl19} used a more direct argument based on the notion of $t$-daisies to find a suitable threshold $m$, \cite{GGS26} used the notion of \emph{spreadness}.
    A distribution over sets $\mu$ is said to be $(m, r)$-spread if, for every non-empty set $T$, the total mass of sets containing $T$ is at most $m/r^{|T|}$.
    Using Janson's inequality, \cite{GGS26} showed that if a distribution is $(m, r)$-spread, then \pref{equation:soundness} holds for $p=\widetilde{\Theta}(1/r)$.
    
    A complementary Lemma of \cite{GGS26} is the \emph{spreadness extraction lemma}, which we also use in this work.
    Roughly speaking, the lemma states that for every distribution over sets of size at most $q$, there exists an $m$ such that after removing the heavy coordinates $H_m$ from every query set and restricting to a large subfamily, the resulting distribution is $(m,n^{\Theta(1/q)})$-spread.
    Together, these two results show that there exists an $m$ for which \pref{equation:soundness} holds with $p = n^{-\Theta \left(1/q\right)}$, yielding the nearly tight lower bound of \cite{GGS26}.

\subsection{The Challenge of Adaptivity} \label{section:adaptive-challange}
    An adaptive relaxed decoder may choose to query coordinates depending on the answers to its previous queries.
    This raises two immediate challenges:
    \begin{itemize}
        \item \textbf{No single distribution over sets.} The spreadness extraction lemma operates on a fixed distribution over query sets.
        It is therefore unclear how to apply it, or any other method formulated for such a distribution, in the adaptive setting.
        \item \textbf{No input-independent notion of ``heavy'' indices.} Since the query distribution depends on the input, so do the degrees of the individual coordinates.
        Therefore, for a threshold $m$, it is unclear which set should play the role of $H_m$.
    \end{itemize}
    To overcome these challenges, we exploit the particular structure imposed by adaptivity.

    \paragraph{Distributions over decision trees. }
    Any adaptive algorithm can be described by sampling from a distribution over \emph{decision trees}, evaluating the sampled tree on the input $y$, and outputting the value at the corresponding leaf.
    Accordingly, any adaptive relaxed decoder with query complexity $q$ can be fully described by a collection of distributions over decision trees, $\{\tau_i\}_{i \in [k]}$, where every tree has depth at most $q$.\footnote{This model also captures $2$-sided error RLDCs.
    Two different trees might give the same query set for some specific input, but reach different output leaves.}

    For any fixed index $i$ and input $y$, each decision tree collapses to a path.
    Hence, for this $y$ the distribution over decision trees induces a distribution over query sets, denoted $\tau^y = \tau_i^y$. This suggests the following tempting but flawed approach.
    
    \paragraph{A naive solution.}
    Consider $\tau^{C(x)}$, the set distribution corresponding to the operation of the relaxed decoder on the correct codeword.
    This is a distribution over query sets, as in the non-adaptive case, so we can attempt to apply the spreadness extraction lemma to obtain a threshold $m$ and a set of heavy indices $H$, and then proceed as in the non-adaptive case.
    
    However, this approach faces two problems:
    \begin{itemize}
        \item \textbf{Unknown input codeword.}
        The sample-based decoder does not know $C(x)$, and hence cannot base its operation on $\tau^{C(x)}$ or on a heavy set extracted from it.
        \item \textbf{Circular dependency.}
        To enumerate assignments $\kappa$ to $H$, the decoder must first determine $H$, but each assignment induces a different input $y$, whose query distribution $\tau^y$ may have a different set of heavy indices.
    \end{itemize}

    \paragraph{Previous approaches. }
    A well-known way to reduce decision trees to sets, which was used in \cite{gl19}, is to query all branches of each decision tree.
    However, doing so can increase the query complexity \emph{exponentially}, from $q$ to $2^q$, leading to a much worse lower bound.
    
    Dall’Agnol, Gur, and Lachish \cite{DGL23} extended the $n = k^{1+\Omega\left(\frac{1}{q^2 \log ^2 q}\right)}$ lower bound of \cite{gl19} to adaptive, 2-sided error RLDCs, while avoiding this exponential query blowup.
    We next describe their approach and the obstacles to obtaining a stronger lower bound within the same framework.
    
    \paragraph{Flattening the tree distribution.}
    The first step in \cite{DGL23} is to replace the distribution over decision trees by a single ``flat'' multiset containing the query set of every branch of every tree.
    This yields an \textbf{input-independent} structure, allowing notions such as coordinate weights to be defined globally and methods for set distributions to be applied.
    
    The cost is that the degree of a coordinate under the distribution $\tau^y$ induced by a particular input $y$ can be larger by a factor of up to $2^q$ than its degree in the flat distribution.
    Hence, to capture every coordinate whose weight may exceed $m$ for some input, the global heavy set must include all coordinates of weight at least $m/2^q$ in the flat distribution.
    This can increase the size of $H$ by a factor of $2^q$, causing a corresponding loss in the resulting lower bound.
    
    For constant $q$, as considered in \cite{DGL23}, this loss is absorbed into the constants.
    In the regime of $q=\omega(1)$, however, this yields a non-constant loss in the lower bound.
    \paragraph{Lossless partitioning.}
    After the previous step, \cite{DGL23} partitions the flat query distribution into $q+1$ structures, called $t$-daisies.
    For every fixed input $y$, at least one of these daisies has mass $\Omega(1/q)$ under $\tau^y$.
    To ensure that this daisy still contains sufficiently more mass on correct local views than on erroneous ones, the soundness error must therefore be $O(1/q)$.

    Accordingly,  as a preprocessing step, \cite{DGL23} performs soundness amplification, reducing the relaxed decoder's soundness error to $O(1/q)$ by repeating it $\Theta(\log q)$ times.
    This results in a corresponding blowup in query complexity and hence a loss in the final lower bound.

    In the non-adaptive setting, techniques developed in \cite{Goldreich23, GGS26} avoid this loss by extracting a single structure whose mass is arbitrarily close to $1$.
    However, \emph{this does not extend to the adaptive setting}: a small fraction discarded from the global flat distribution may contain all query sets active on some specific input $y$.
    Thus, within the framework of \cite{DGL23}, the lossless partitioning, the soundness amplification and the resulting $O(\log q)$ query-complexity blowup appear unavoidable.

    \paragraph{The limitations of consensus.}
    The above partitioning raises yet another difficulty.
    Checking consensus separately within each part is insufficient, since all good active paths for a specific input may belong to other parts.
    On the other hand, checking consensus across all parts simultaneously requires guessing a common assignment to their heavy indices, so that all simulated paths correspond to the same input.
    The resulting heavy set may be too large, requiring a soundness error probability smaller than what the analysis can guarantee.

    \cite{DGL23} addresses this difficulty by replacing consensus with voting.
    For each assignment, the sampled local views cast votes, and an output is declared only if its number of votes crosses a threshold $\theta$.
    However, this strategy introduces a new challenge in the soundness analysis.
    
    \paragraph{Upper-tail concentration.}
    As discussed above, for any fixed guess, the probability that the sample-based decoder outputs an incorrect bit must be exponentially small in the number of heavy indices.
    With the consensus rule, this is done by identifying, for each wrong guess, a family of good sets, and showing that, except with exponentially small probability, at least one of them is contained in $W \cup H$.
    Equivalently, if $X$ denotes the number of captured good sets, we seek a bound of the form
    \[
        \Pr[X = 0] \leq \exp(-\Omega(|H|)) \; .    
    \]
    While \cite{gl19} establishes this bound using a large family of disjoint sets, \cite{GGS26} extracts a spread distribution and applies Janson's inequality, allowing a smaller sampling probability.

    With voting, soundness instead requires an upper-tail bound of the form
    \[
        \Pr[Y \geq \theta] \leq \exp(-\Omega(|H|)) \;,
    \]
    where $Y$ is the number of votes for an incorrect output. 
    To prove this bound, \cite{DGL23} uses the Hajnal--Szemerédi theorem to obtain disjoint sets with independent sampling events, allowing Chernoff bounds for both tails.

    One might expect that the spreadness extraction and Janson's inequality used in \cite{GGS26} would yield the required upper-tail bound while retaining the smaller sampling probability.
    However, \textbf{Janson's inequality provides exponential lower-tail concentration but no symmetric upper-tail bound}.
    This is not merely a limitation of Janson's inequality.
    In contrast to disjointness, spreadness alone does not imply a corresponding exponential \emph{upper-tail} bound, as we demonstrate next.
    
    \paragraph{Spreadness does not imply exponential upper-tail concentration. }
    We construct an $(m,r)$-spread distribution for which $\Pr[X \geq 2\E[X]]$ is much larger than $\exp(-\Omega(1/m))$, where $X$ is the number of sampled sets for $p=1/r$.
    Such a bound is required for soundness, since the heavy set $H_m$ may have size $\Theta(q/m)$.
    
    Consider $r$ vertex-disjoint copies of the clique $K_r$, and let $\mu$ be the uniform distribution over their edges.
    This is a set family of $r^2$ base elements and $r {r \choose 2} = \frac{r^2 (r-1)}{2}$ sets, and the mass of each set is $\frac{2}{r^2 (r-1)}$.
    It is straightforward to verify that $\mu$ is $(m,r)$-spread for $m = \frac{2}{r-1}$.
    The probability of sampling each fixed set is $p^2 = r^{-2}$, and hence $2 \E[X] = \frac{2}{r^2} \cdot \frac{r^2(r-1)}{2} = r-1$.
    
    Fix a clique and $2 \sqrt{r}$ elements in it, and consider the event that all of these elements are sampled.
    On this event, $X \geq {2 \sqrt{r} \choose 2} \approx 2r > 2 \E[X]$.
    On the other hand, the probability of this event is $\left(\frac{1}{r}\right)^{2 \sqrt{r}} \approx \exp(-2 \sqrt{r} \log r)$.
    Hence, $\Pr[X \geq 2 \E[X]] \geq  \exp(-2 \sqrt{r} \log r)$, which is much larger than the required $\exp(-1/m) = \exp(-\Theta(r))$.\footnote{We remark that this cannot be fixed by increasing $p$ by a small factor, such as $\log r$.
    In fact, increasing $p$ from $\frac{1}{r}$ does not remove this obstacle until $p = \Theta(\frac{1}{\sqrt{r} \log r})$.}

    \paragraph{The challenge of two-sided error. }
    The upper-tail concentration obstacle also arises for non-adaptive decoders with two-sided error.
    In this setting, voting is the natural replacement for consensus, which again requires an exponentially small probability that the number of incorrect votes exceeds the threshold.
    As discussed above, spreadness alone does not provide such a bound.

    \paragraph{Two-sided error in property testing.}
    Carenini \cite{Carenini26} solves the two-sided error challenge for \emph{non-adaptive property testers}. 
    Building on \cite{DGL23}, she shows a reduction from query-based to sample-based testing with near-optimal overhead, implicitly using what can be viewed as the (non-capped) HT estimator and a new ranked spreadness extraction lemma instead of a partition into disjoint sets.
    
    However, her approach does not directly extend to RLDCs.
    Unlike RLDCs, property testers have only two possible outputs, so a low rejection probability is equivalent to a high acceptance probability.
    This allows Carenini to reverse the threshold test of \cite{DGL23}: her algorithm accepts if some assignment to the heavy coordinates yields an estimated rejection probability \emph{below} a threshold.
    In the soundness case, every assignment has a high rejection probability, so ruling out false acceptance requires exponential lower-tail concentration.

    In contrast, for RLDCs, a low probability of outputting a bit $b$ does not imply a high probability of outputting $1-b$, since the decoder may output $\bot$.
    Thus, within this voting framework, the sample-based decoder must instead output a bit only if its estimated probability exceeds a threshold, requiring exponential upper-tail concentration.

\subsection{Our Approach}
    We overcome the limitations above using the following ideas:
    \begin{enumerate}
        \item \textbf{Guessing the heavy indices: } 
        Instead of finding a global set of heavy coordinates, we enumerate all candidate sets $H$ that could be obtained by applying the spreadness extraction lemma, together with all possible assignments to them.
        
        Crucially, our soundness analysis does not require spreadness for any of these guesses, and provides a sufficiently strong bound to union bound over all of them.
        \item \textbf{Decoding via mass estimation: }
        Instead of using a consensus or voting mechanism, we decode by estimating the probability that the relaxed decoder outputs each bit. 
        We then reduce this task to estimating the total mass of an arbitrary sub-distribution.
        \item \textbf{The Capped Horvitz--Thompson Estimator: }
        We develop the CHT estimator, a downward-biased estimator for the total mass of a sub-distribution.
        \begin{enumerate}
            \item After normalization by the cap, the CHT estimator is strongly self-bounding, which implies exponential concentration around its mean.
            This property is sufficient for the soundness analysis.
            \item We prove that if the estimated sub-distribution is spread, then the bias introduced by capping is small.
            \item For completeness, we apply the spreadness extraction lemma to the correct-output sub-distribution induced by the correct codeword.
            The extracted sub-distribution retains sufficient mass and satisfies spreadness, ensuring that the expected CHT estimate is large enough.
        \end{enumerate}
    \end{enumerate}

    Our solution exhibits an \textbf{asymmetry between soundness and completeness}: we do not need accurate estimates for every candidate.
    For soundness, the downward bias of CHT is beneficial, and concentration holds regardless of spreadness, allowing us to handle all guesses uniformly even when they induce different query distributions.
    For completeness, accuracy is required only for the correct guess, where spreadness guarantees that the capping loss is small.

    We remark that we use spreadness in a substantially different way from \cite{GGS26}.
    While \cite{GGS26} uses spreadness to apply Janson's inequality, we use it to directly bound the variance of the discrete derivatives of the Horvitz--Thompson estimator, thereby controlling the bias introduced by capping.

    We next describe these ideas in more detail and give a high-level overview of the proof of the main theorem.
    For simplicity, throughout the overview we track the exponent of the lower bound only up to $1+\Theta(1/q)$ and do not optimize the constant hidden in the $\Theta$ notation.
    
    \paragraph{Our sample-based decoder. }
    Recall that the sample-based decoder receives a valid codeword $c = C(x)$.
    It samples a set $W \subseteq [n]$ by including each coordinate independently with probability $p$, and reads $c|_W$.
    Specifically, we choose
    \[
        p = \frac{q^2 \log n}{n^{\Theta(1/q)}} \; ,
    \]
    which directly translates to the nearly-tight lower bound.

    The sample-based decoder then uses the same sample $(W, c|_W)$ to decode all of $x$.
    To decode $x_i$, it iterates over the $O(q)$ possible thresholds $m$ prescribed by the spreadness extraction lemma, all candidate sets $H$ of the corresponding size $|H|=\lfloor q/m\rfloor$, and all assignments $\kappa \in \{0,1\}^H$.

    In each iteration, the decoder considers a hypothetical input string $y$, obtained from the correct codeword $c$ by replacing its values at the coordinates in $H$ with those specified by $\kappa$.
    We denote the resulting string by $y=c_{H\leftarrow\kappa}$.
    By our choice of parameters, $|H| \leq \delta n$, where $\delta$ is the decoding radius, so $y$ is within the decoding radius of $c$.
    
    For each output bit $b \in \{0, 1\}$, it then estimates the probability that the relaxed decoder outputs $b$ on this string, namely
    \[
        \Pr_{T \sim \tau}[T(y)=b] \;,
    \]
    where $T(y)$ denotes the output of the decision tree $T$ on input $y$, and $\tau=\tau_i$ is the distribution over decision trees corresponding to index $i$.
    If the estimate is at least $1/2$, the sample-based decoder outputs $b$ as the value of $x_i$ and halts.
    Otherwise, it continues to the next guess.

    This algorithm raises the obvious question:
    how can the sample-based decoder estimate $\Pr[T(y)=b]$ if it does not know all of $y$?

    \paragraph{Reduction to mass estimation. }
    The key insight is that for every fixed $H, \kappa$, and $b$, we can decompose this probability according to the set of queried indices outside $H$.
    \[
        \Pr_{T \sim \tau}[T(y)=b] = \sum_{S \subseteq [n] \setminus H} \Pr_{T \sim \tau}[T(y) = b \text{ and } Q_T(y) \setminus H = S] \; ,
    \]
    where $Q_T(y) \subseteq [n]$ denotes the coordinates queried by $T$ for the input $y$.
    The events in the sum are disjoint, and their union is exactly the event $T(y)=b$. 

    We define the sub-distribution $\rho_{H,\kappa,b}(S) = \Pr_{T \sim \tau}[T(y) = b \text{ and } Q_T(y) \setminus H = S]$.
    Then, the probability we seek to estimate is exactly the total mass of this sub-distribution:
    \[
        \Pr_{T \sim \tau}[T(y)=b] = \sum_{S \subseteq [n] \setminus H}{\rho_{H,\kappa,b}(S)} = \|\rho_{H,\kappa,b}\|_1 \; .
    \]
    
    This perspective is useful because, for every fixed $H$ and assignment $\kappa$ (and the induced string $y$), the sample-based decoder can compute $\rho_{H,\kappa,b}(S)$ for every $S \subseteq W$.

    To compute $\rho_{H,\kappa,b}(S)$, the decoder iterates over all trees $T \in \supp(\tau)$ and simulates their execution on $y$.
    Queries to indices in $H$ are answered according to $\kappa$, and queries to indices in $W \setminus H$ are answered using the sampled view.
    If, during the simulation, $T$ queries an index outside $H \cup W$, the simulation may be stopped, since in this case $Q_T(y) \setminus H \neq S$.
    Otherwise, once the simulation is complete, if $T(y)=b$ and $Q_T(y)\setminus H=S$, the decoder adds $\tau(T)$ to a running sum.
    After iterating over all trees, this sum is exactly $\rho_{H,\kappa,b}(S)$.

    The decomposition points us to the following concrete goal:
    \begin{center}
        Estimating $\|\rho_{H,\kappa,b}\|_1$ from the values $\rho_{H,\kappa,b}(S)$ for sets $S \subseteq W$.
    \end{center}

    \subsection{The Capped Horvitz--Thompson estimator}
    As discussed in \pref{section:intro-cht}, a natural tool for the foregoing goal is the Horvitz--Thompson (HT) estimator. Although unbiased, the HT estimator need not concentrate around its mean.
    To overcome this issue, we introduce the \emph{Capped Horvitz--Thompson} estimator.

    We design the CHT estimator so that, after normalization by the cap $\Gamma$, it is a \emph{strongly self-bounding} function; 
    concentration inequalities for self-bounding functions due to Boucheron, Lugosi, and Massart \cite{BLM09} (see also \cite[Section~6.11]{BLM13}) then imply exponential concentration around its mean.

    Roughly speaking, a function $f$ is \emph{strongly $(a, b)$-self-bounding} if (1) for every $i$ and $x$, its $i$-th marginal contribution is bounded by $1$, namely $0 \leq f(x) - f(x^{(i \to 0)}) \leq 1$, where $x^{(i \to 0)}$ denotes the vector obtained from $x$ by setting its $i$-th coordinate to $0$; and (2) for every $x$, the sum of its marginal contributions is bounded in terms of the value of the function itself:
    \[
        \sum_{i \in [n]}{ ( f(x) - f(x^{(i \to 0)} ))} \leq af(x) + b \; .
    \]
    
    For a sub-distribution $\rho$ and a fixed $x$, let $\overline{R}_{\rho,\Gamma}(x) = \sum_S w_S$ where $\{w_S\}$ are optimal weights realizing the maximum in the definition of the $\Gamma$-capped HT estimator (CHT; see Definition~\ref{def:CHTintro}).
    Its marginal contributions satisfy
    \[
        \overline{R}_{\rho,\Gamma}(x) - \overline{R}_{\rho,\Gamma}(x^{(i \to 0)}) \leq \sum_{S \ni i}w_S \leq \Gamma \; ,
    \]
    where the second inequality holds by the definition of the cap.
    
    Hence, since setting a coordinate to $0$ cannot increase the CHT estimate, the normalized CHT estimator $f(x)=\frac{\overline{R}_{\rho,\Gamma}(x)}{\Gamma}$ satisfies the first requirement.
    The second requirement also holds, with $a=q$ and $b=0$:
    by changing the summation order, and since $\rho$ is supported on sets of size at most $q$,
    \[
        \sum_i{(\overline{R}_{\rho,\Gamma}(x) - \overline{R}_{\rho,\Gamma}(x^{(i \to 0)}))}
        \leq \sum_i \sum_{S \ni i}w_S
        = \sum_S |S|w_S
        \leq q \sum_S w_S = q \overline{R}_{\rho,\Gamma}(x) \; .
    \]
    Normalizing by $\Gamma$ shows that $f$ satisfies the second requirement as well.

    For the soundness analysis, we apply the upper-tail concentration inequality for self-bounding functions \cite[Theorem~6.19]{BLM13} to $f$ and deduce that if $\E[\overline{R}_{\rho, \Gamma}(X)] \leq \frac12 - \Omega(1)$, then
    \[
        \Pr[\overline{R}_{\rho, \Gamma}(X) > \frac{1}{2}] = \exp(-\Omega(\frac{1}{q \Gamma})) \;.
    \]
    Remarkably, this inequality does not rely on any structural properties of the estimated sub-distribution.

    Similarly, for the completeness analysis, if $\E[\overline{R}_{\rho, \Gamma}(X)] \geq \frac12 + \Omega(1)$, then the lower-tail concentration inequality for self-bounding functions \cite[Theorem~6.20]{BLM13} gives
    \[
        \Pr[\overline{R}_{\rho, \Gamma}(X) \leq \frac{1}{2}] = \exp(-\Omega(\frac{1}{q \Gamma})) \;.
    \]
    
    \paragraph{Soundness analysis. }
    For brevity, let us denote $\rho = \rho_{H,\kappa,b}$.
    Let us also denote by $R_\rho, \overline{R}_\rho$ the HT estimator and the capped HT estimator, respectively, for the sub-distribution $\rho$.
    Since $\overline{R}_\rho$ is pointwise upper-bounded by the unbiased Horvitz--Thompson estimator, we have
    \[
        \E[\overline{R}_\rho(X)] \leq \E[R_\rho(X)] = \| \rho \|_1 \;.
    \]

    Recall that $y$ is within the decoding radius of the RLDC.
    Hence, by the soundness guarantee of the relaxed decoder, if $b$ is the incorrect output bit, then
    \[
        \| \rho \|_1  = \Pr_{T \sim \tau}[T(y)=b] \leq 1/3 \;,
    \]
    and therefore $\E[\overline{R}_\rho(X)] \leq 1/3$.
    
    Thus, by choosing a sufficiently small cap $\Gamma$, and using the upper-tail concentration inequality above, we can make the probability of $\overline{R}_\rho(X) > \frac{1}{2}$ small enough to survive a union bound over all guesses.
    Specifically, we choose $\Gamma=\Theta(\frac{m}{q^2 \log n})$, and obtain an error probability of $\exp(-\Omega(\frac{q \log n}{m}))$ for every fixed guess.
    
    On the other hand, for a fixed $m$ there are $\exp(O(\frac{q \log n}{m}))$ possible guesses of a set $H$ of size $\frac{q}{m}$ and an assignment $\kappa$: there are
    \[
        {n \choose \lfloor \frac{q}{m} \rfloor} \leq n^{q/m} = O(2^{\frac{q}{m} \log n})
    \]
    possible sets, each with $2^{q/m}$ possible assignments.
    By choosing the constants appropriately, the concentration bound is strong enough to take a union bound over all these pairs.

    Finally, and crucially, there are only $\Theta(q)$ possible values of the threshold $m$.
    Taking one additional union bound over these values implies that, with constant probability of, say, $5/6$, no iteration of the sample-based decoder outputs the incorrect bit.
    This completes the soundness analysis.

    \paragraph{Spreadness implies small capping loss}
    The key claim for the completeness argument is that if an arbitrary sub-distribution $\sigma$ is spread, and $\Gamma$ is chosen appropriately, then $\E[\overline{R}_\sigma(X)]$ is close to $\E[R_\sigma(X)]$.
    Consider the marginal contributions of $R_\sigma$,
    \[
        d_i(x) =\sum_{S \in \langle i \rangle}{\sigma(S) p^{-|S|}x_S} \;.
    \]
    Like the HT estimator itself, these functions are unbiased: $\E[d_i(X)] = \sigma(\langle i \rangle)$.
    However, they need not concentrate around their expectations. 
    Their variance may be large, causing $d_i(X)$ to occasionally exceed the cap by a large amount, in which case capping may incur a significant loss.

    We show that if $\sigma$ is spread, then the variance of each $d_i$, conditioned on $i$ being sampled, is small.
    Specifically, we prove that if $\sigma$ is $(m, r)$-spread, then for $\Gamma=2m/pr$ the expected total overload (i.e., the sum, over all indices, of the amount by which $d_i(X)$ exceeds the cap) is at most $O\left(\frac{q^2}{pr}\right)$.
    Thus, if $pr=\omega(q^2)$, this gap can be made arbitrarily small.
    We then show that the gap $\E[R_\sigma(X)] - \E[\overline{R}_\sigma(X)]$ is bounded by this expected total overload.
    
    \paragraph{Completeness analysis. }
    Our goal is to show there exists an iteration, i.e., a threshold $m$, a set $H$, and assignment $\kappa$, for which the sample-based decoder outputs the correct bit $b^* = x_i$ with high probability.
    Denoting $\rho = \rho_{H,\kappa,b^*}$, we want to show $\overline{R}_{\rho}(X) \geq 1/2$ with probability at least $5/6$.
    
    To this end, we define the sub-distribution 
    \[
        \nu(S) = \Pr_{T \sim \tau}[T(C(x)) = b^* \land Q_T(C(x)) = S] \; .
    \]
    By the completeness guarantee of the relaxed decoder, $\| \nu \|_1 = \Pr_{T \sim \tau}[T(C(x)) = b^*] \geq 2/3$.

    We normalize $\nu$ and apply the spreadness extraction lemma \cite[Lemma~5.4]{GGS26}.
    We obtain a threshold $m^*$, a set $H^*$ with $|H^*| = \lfloor \frac{q}{m^*} \rfloor$, and a sub-distribution $\sigma \leq \nu^{\circ H^*}$ that is $(m^*, r)$-spread and satisfies $\|\sigma \|_1 \geq 0.99 \| \nu \|_1 \geq 0.99 \cdot \frac23$.

    Next, we consider the sub-distribution used by the sample-based decoder for the correct guess, $\rho =\rho_{H^*, \kappa^*, b^*}$, where $\kappa^* = C(x)|_{H^*}$.
    By definition,
    \[
        \rho(S) = \Pr_{T \sim \tau}[T(C(x)) = b^* \land Q_T(C(x)) \setminus H^* = S] \; .
    \]
    This is exactly the sub-distribution obtained by puncturing $\nu$ by $H^*$.
    That is, $\rho = \nu^{\circ H^*}$.

    We then apply the key claim above to the spread sub-distribution $\sigma$, and deduce that $\E[R_\sigma(X)] - \E[\overline{R}_\sigma(X)] = o(1)$, and hence $\E[\overline{R}_\sigma(X)] \geq 0.99 \cdot \frac23 - o(1)$.
    Since $\rho = \nu^{\circ H^*} \geq \sigma$, and since the CHT estimator is monotone, we also obtain $\E[\overline{R}_\rho(X)] \geq 0.99 \cdot\frac23 - o(1)$.
    In particular, for sufficiently large $n$, this expectation is bounded away from $1/2$ by a constant.
    
    Finally, we apply the lower-tail concentration inequality for self-bounding functions \cite[Theorem~6.20]{BLM13} and deduce that $\Pr[\overline{R}_\rho(X) \leq 1/2]$ is sufficiently small, which completes the completeness proof.

    \paragraph{Putting everything together. }
        Combining the soundness and completeness analyses, the sample-based decoder recovers each message bit with probability at least $2/3$ while sampling every coordinate independently with probability $p = \frac{q^2 \log n}{n^{\Theta(1/q)}}$, yielding the stated lower bound.

\section{Preliminaries}
    We provide notation used throughout the paper, together with known facts and theorems.
    
\subsection{Basic notation and facts}
    \paragraph{Set notation.} Let $U$ be a finite set.
    \begin{itemize} 
        \item For a set $U$, we denote by $\mathcal{P}(U)$ the power set of $U$, i.e., the family of all subsets of $U$. Furthermore, let $\mathcal{P}_{\leq q}(U) = \{S \in \mathcal{P}(U) \;;\; |S| \leq q\}$ denote the family of subsets of size at most $q$. 
    
        \item The \emph{star} of a set $T \subseteq U$, denoted by $\langle T \rangle$, is the family of subsets of $U$ that contain $T$;
        that is, $\langle T \rangle = \{ S \subseteq U \mid S \supseteq T\} \subseteq \mathcal{P}(U)$.

        We use a slight abuse of notation and write $\langle x \rangle$ to denote $\langle \{ x \} \rangle$ for an element $x \in U$.

    \end{itemize}

    \paragraph{Sub-distributions.}
    Let $D$ be a discrete domain.
    
    \begin{itemize}
        \item A \emph{sub-distribution} $\mu$ over $D$ is a function $\mu: D \rightarrow [0, 1]$ such that $\sum_{x \in D}\mu(x) \leq1$.
        If $\sum_{x \in D}\mu(x) = 1$, we say that $\mu$ is a (probability) \emph{distribution}.

        \item For any subset $A \subseteq D$, the \emph{mass} of $A$ is $\mu(A) = \sum_{x \in A}\mu(x)$.

        \item The \emph{total mass} of a sub-distribution $\mu$ is $\|\mu \|_1 =\sum_{x \in D}\mu(x)$.

        \item The \emph{support} of a sub-distribution $\mu$ is the set of elements with non-zero probability, denoted  $\supp(\mu)=\{ x \in D \mid \mu(x) > 0\}$.

        \item The \emph{conditioning} of a sub-distribution $\mu$ on a set $A \subseteq D$ with positive mass is a distribution over $A$, denoted $\mu_A$, defined for each $x \in A$ by $\mu_A(x)=\frac{\mu(x)}{\mu(A)}$.

        \item The \emph{restriction} of a sub-distribution $\mu$ to a set $A \subseteq D$ is a sub-distribution over $D$, denoted $\mu|_A$, defined for each $x \in D$ by
        \[ 
            (\mu|_A)(x)= \begin{cases}
                            \mu(x)&x\in A,\\
                            0 & x \notin A.
                            \end{cases}
        \]
        \begin{fact}
            For every sub-distribution $\mu$ and a set $A \subseteq D$ with positive mass, $\mu|_A(x) = \mu(A) \mu_A(x)$ for every $x \in A$.
        \end{fact}

    \end{itemize}

\paragraph{Additional notations and facts. }
    \begin{itemize}
        \item For a value $a \in \R$, we denote $a_+ = \max(a,0)$.
        \item We denote $[n] = \{1, 2, 3, \ldots, n\}$.
        \item All of the logarithms in this paper are in base $2$.
    \end{itemize}
    
\begin{fact} \label{fact:geometric-sum}
    For every $b > a > 0$,
    \[
        \sum_{i=1}^\infty \left (\frac{a}{b}\right)^i = \frac{a}{b-a} \; .
    \]
\end{fact}

\subsection{Information Theory and Entropy}
\begin{definition}
    For $t \in (0, 1)$, the binary entropy function is
    \[
        h_2(t)=-t\log t-(1 - t) \log(1-t) \; .
    \]
\end{definition}

\begin{definition}
    The \emph{entropy} of a discrete random variable $X$ is defined by
    \[
        H(X)=-\sum_x \Pr[X=x]\log \Pr[X=x] \; .
    \]
\end{definition}

\begin{definition}
    For random variables $X$ and $Y$, the conditional entropy of $X$ given $Y$ is
    \[
        H(X\mid Y) = \sum_y \Pr[Y=y] H(X\mid Y=y) \; .
    \]
\end{definition}

\begin{fact} [Fano's inequality for binary random variables \cite{Fano61}] \label{fact:fano}
    Let $X \to Y \to \widehat{X}$ be a Markov chain, where $X$ and $\widehat{X}$ are binary random variables.
    Let $\varepsilon = \Pr[\widehat{X} \neq X]$.
    Then $H(X \mid Y) \leq h_2(\varepsilon)$.
\end{fact}

\begin{fact}[Chain rule for conditional entropy] \label{fact:chain}
For random variables $X,Y,Z$,
\begin{align*}
    H(X,Y \mid Z)
    &= H(X \mid Z) + H(Y \mid X,Z) \\
    &= H(Y \mid Z) + H(X \mid Y,Z) \; .
\end{align*}
    In particular, if $Y$ is determined by $X$ and $Z$, then $H(Y \mid X,Z) = 0$, and
    \[
        H(X \mid Y,Z) = H(X \mid Z) - H(Y \mid Z) \; .
    \]
\end{fact}

\subsection{Spreadness}
\begin{definition}[$(m, r)$-spread distributions] \label{def:spread-distribution} 
    Let $\mu$ be a sub-distribution over $\mathcal{P}(U)$ for a finite universe $U$. Let $r > 1$ and let $m > 0$.
    We say that $\mu$ is \emph{$(m, r)$-spread} if for any non-empty set $S \subseteq U$,
    \[
        \mu(\langle S \rangle) \leq \frac{m}{r^{|S|}} \; .
    \]
\end{definition}

\begin{definition}[Punctured Distribution] \label{defn:punctured_distribution}
    Let $\mu$ be a sub-distribution over $\mathcal{P}(U)$.
    The \emph{punctured distribution} of $\mu$ with respect to a set $H \subseteq U$ is a sub-distribution over $\mathcal{P}(U \setminus H)$, denoted $\mu^{\circ H}$, defined, for every $A \subseteq U\setminus H$, by
    \[
        \mu^{\circ H}(A) = \sum_{B \subseteq H} \mu(A \cup B) \; .
    \]

    Equivalently, puncturing maps each set $S$ to $S\setminus H$ while preserving its weight.
    
    Note that
    \[
        \sum_{A \in \mathcal{P}(U \setminus H)} \mu^{\circ H}(A) = \|\mu\|_1 \; .
    \]
    Hence, puncturing preserves the total mass, and in particular, if $\mu$ is a distribution, then $\mu^{\circ H}$ is also a distribution.
\end{definition}

\begin{lemma} (The Spreadness Extraction Lemma, Lemma 5.4 in \cite{GGS26}) \label{lemma:large_spread_daisy}
    Let $\mu$ be a distribution over $\mathcal{P}_{\leq q}(U)$, where $|U| = n$.
    Let $d > q$ be any integer.
    
    Then, there exists $j \in [d]$, a family of sets $\D \subseteq \mathrm{supp}(\mu)$, and a non-empty set of elements $H \subseteq U$ such that the distribution $(\mu_\D)^{\circ H}$ is $(m, n^{1/d})$-spread, where:
    \begin{align*}        
        m &= \frac{q}{\mu(\D)} \cdot \frac{n^{j/d}}{n}, &
        |H| & \leq \frac{n}{n^{j/d}}, &
        \mu(\D) &\geq 1 - \frac{q}{d} \;.
    \end{align*}
\end{lemma}

\subsection{Relaxed Locally Decodable Codes} \label{section:ecc}
    We recall the formal definition of relaxed locally decodable codes \cite{BGHSV06}.

    \begin{definition} (Relaxed locally decodable codes) \label{def:rldc}
        Let $C:\{0,1\}^k \rightarrow \{0, 1\}^n$ be an error correcting code.
        A $(q, \delta,\alpha,\beta)$-relaxed decoder for $C$ is a randomized procedure $\B$ that on an explicit input $i \in [k]$, and oracle access to $w \in \{0, 1\}^n$, outputs an element of $\{0, 1, \bot\}$, and satisfies the following requirements: 
        \begin{enumerate}
            \item (completeness) If $w = C(x)$ for some $x \in \{0, 1\}^k$, then $\Pr[\B^{w}(i) = x_i] \geq \alpha$.
            \item (relaxed local decoding) If there exists $x \in \{0, 1\}^k$ such that $\dist(w, C(x)) \leq \delta$, then $\Pr[\B^{w}(i) \in \{x_i, \bot\}] \geq \beta$.
            \item For every input $i$ and oracle access to any $w \in \{0, 1\}^n$, $\B$ makes at most $q$ queries.
        \end{enumerate}
        We say that $\B$ is non-adaptive if it determines all its queries based on its explicit input (namely, the index to decode) and internal coin tosses, independently of the specific $w$ to which it is given oracle access.
        
        We refer to $\delta$ as the decoding radius of the decoder, to $\alpha$ as its completeness probability, and to $\beta$ as its soundness probability.
        
        If $\alpha=1$, we say that the relaxed decoder has $1$-sided error.
        Unless stated otherwise, we assume throughout the paper that $\alpha=\beta=2/3$.
    \end{definition}

\section{The Capped Horvitz--Thompson Estimator}
In this section we present and prove the main tools we need for constructing the sample-based decoder.
In \pref{section:ht}, we define the Horvitz--Thompson (HT) Estimator, and present its basic properties.
In \pref{section:cht} we define the \emph{Capped} Horvitz--Thompson (CHT) Estimator, and compare it to the HT estimator.

We then study the concentration and expectation of the CHT estimator.
Specifically, in \pref{section:self-bounding} we prove that the CHT estimator is \emph{strongly self-bounding}, and in \pref{section:concentration} we use this and the BLM inequalities to prove that the CHT concentrates around its mean.
In \pref{section:expectation}, we prove that if the underlying sub-distribution is \emph{spread}, then the expectation of the CHT estimator is not much smaller than that of the HT estimator.

\subsection{The Horvitz--Thompson Estimator} \label{section:ht}

    Our starting point is the problem of estimating the total mass of a sub-distribution $\mu$ from a set of coordinates, each sampled independently with probability $p$.
    A set $S$ is then fully sampled with probability $p^{|S|}$, so weighting its contribution by $p^{-|S|}$ compensates for the sampling and yields an unbiased estimator.
    This motivates the Horvitz–Thompson estimator defined below. 
\begin{definition} [Horvitz--Thompson (HT) Estimator]
    Let $\mu$ be a sub-distribution over subsets of $[n]$ and let $p \in (0, 1]$.
    That is, for every $S \subseteq [n]$, $\mu(S) \geq 0$ and $\|\mu \|_1 = \sum_S{\mu(S)} \leq 1$.
    
    The \emph{$p$-Horvitz--Thompson Estimator} for $\mu$ is a function $R_{\mu, p}: \{0, 1\}^n \rightarrow \R$ defined by
    \[
            R_{\mu,p}(x) = \sum_{S \subseteq[n]} \mu(S) p^{-|S|} x_S \; ,
    \]
    where $x_S = \prod_{i \in S}{x_i}$. When it is clear from context, we will suppress $p$ and write $R_{\mu}(x)$.
\end{definition}

A basic fact about the HT estimator, which gives the motivation to use it, is that it is an unbiased estimator of mass $\|\mu\|_1$.
\begin{observation} [The HT estimator is unbiased]
    If $X =(X_1,\ldots , X_n)$ is a vector of independent Bernoulli random variables with probability $p$, then $\E[X_S] = p^{|S|}$ and the expectation of the Horvitz--Thompson Estimator is
    \[
        \E[R_\mu(X)] = \sum_{S \subseteq[n]} \mu(S) p^{-|S|} \E[X_S] = \sum_{S \subseteq[n]} \mu(S) \; .
    \]
\end{observation}

An important aspect of the HT estimator that we study are its \emph{discrete derivatives}.
\begin{definition} [Discrete Derivative]
    Let $f: \{0, 1\}^n \rightarrow \R$.
    For an index $i \in [n]$, define the \emph{discrete derivative} $\Delta_i f: \{0, 1\}^{n-1} \rightarrow \R$ of $f$ with respect to index $i$ as:
    \[
         \Delta_i f(x^{(i)}) = f(x^{(i \to 1)}) - f(x^{(i \to 0)}) \; ,
    \]
    where $x^{(i \to b)}$ denotes the vector $x$ with its $i$-th coordinate fixed to $b$, and  \\ $x^{(i)} = (x_1, \ldots, x_{i-1}, x_{i+1}, \ldots, x_n)$.
\end{definition}

\begin{claim} [The discrete derivative of HT estimator] \label{claim:cht-derivative}
    Let $\mu$ be a sub-distribution over subsets of $[n]$.
    Let $R_\mu$ be the $p$-Horvitz--Thompson Estimator for $\mu$ for $p \in (0, 1]$.
    Then for every $i \in [n]$,
    \[
        \Delta_i R_\mu(x^{(i)}) = \sum_{S \in \langle i \rangle}\mu(S) p^{-|S|} x_{S \setminus \{i \}} \; ,
    \]
     where $\langle i \rangle = \{ S \subseteq [n] \mid i \in S\}$.
\end{claim}
Note that a similar equality holds for the discrete derivative of any multilinear polynomial.
\begin{proof}
    We express the HT estimator for $\mu$ as
    \[
         R_\mu(x) = \sum_{S \subseteq[n]} \mu(S) p^{-|S|} x_S 
         = \sum_{S \notin \langle i \rangle}\mu(S) p^{-|S|} x_S + x_i\sum_{S \in \langle i \rangle}\mu(S) p^{-|S|} x_{S \setminus \{i\}} \; .
    \]
    Now:
    \[
        R_\mu(x^{(i \to 0)}) = \sum_{S \notin \langle i \rangle}\mu(S) p^{-|S|} x_S + 0 \cdot \sum_{S \in \langle i \rangle}\mu(S) p^{-|S|} x_{S \setminus \{i\}} \;,
    \]
    and
    \[
        R_\mu(x^{(i \to 1)}) = \sum_{S \notin \langle i \rangle}\mu(S) p^{-|S|} x_S + 1 \cdot \sum_{S \in \langle i \rangle}\mu(S) p^{-|S|} x_{S \setminus \{i\}} \; .
    \]
    Hence:
    \[
        \Delta_i R_\mu(x^{(i)}) = R_\mu(x^{(i \to 1)}) - R_\mu(x^{(i \to 0)}) =  \sum_{S \in \langle i \rangle}\mu(S) p^{-|S|} x_{S \setminus \{i\}} \; .
    \]
\end{proof}

\subsection{The Capped Estimator} \label{section:cht}
    While the Horvitz--Thompson estimator is an unbiased estimator, it does not necessarily concentrate around its mean. 
    This motivates capping the Horvitz--Thompson estimator (or more precisely, its discrete derivatives) to ensure concentration.
\begin{definition} [Capped Horvitz--Thompson (CHT) Estimator]
    Let $\mu$ be a sub-distribution over subsets of $[n]$.
    Fix a cap $\Gamma > 0$ and $p \in (0, 1]$.
    
    The \emph{$\Gamma$-Capped $p$-Horvitz--Thompson Estimator} for $\mu$ is the function $\overline{R}_\mu: \{0, 1\}^n \rightarrow \R$ defined by:
    \[
        \overline{R}_{\mu,p,\Gamma}(x) = \max_{\{w_S\}} \sum_{S \subseteq[n]}{w_S} \; ,
    \]
    where the maximum is subject to:
    \[
        0 \leq w_S \leq  \mu(S) p^{-|S|} x_S \; ,
    \]
    and for every $i \in [n]$:
    \[
        \sum_{S \in \langle i \rangle} w_S \leq \Gamma \; .
    \]
    When it is clear from context, we will suppress $p$ and $\Gamma$, and write $\overline{R}_{\mu}(x)$.
\end{definition}

We next state and prove a few immediate properties of the CHT estimator.
\begin{claim} \label{claim:cht-properties}
    Fix $p\in(0,1]$ and $\Gamma>0$, and let $\rho,\rho'$ be
    sub-distributions over subsets of $[n]$.
    Write $R_\sigma=R_{\sigma,p}$ and $\overline{R}_\sigma=\overline{R}_{\sigma,p,\Gamma}$ for $\sigma\in\{\rho,\rho'\}$.
    The following hold for every $x,y\in\{0,1\}^n$ and
    every $i\in[n]$.
    \begin{enumerate}
        \item $\overline{R}_\rho(x) \leq R_\rho(x)$.
        \item For every random variable $X$ over $\{0,1\}^n$, $\E[\overline{R}_\rho(X)] \leq \E[R_\rho(X)]$.
        \item If $x \leq y$ pointwise, then $\overline{R}_\rho(x) \leq \overline{R}_\rho(y)$.
        \item $0 \leq \Delta_i \overline{R}_\rho (x^{(i)}) \leq \Gamma$.
        \item If $\rho \leq \rho'$ pointwise, then $\overline{R}_\rho(x) \leq \overline{R}_{\rho'}(x)$.
    \end{enumerate}
\end{claim}

\begin{proof}
    Let $w_S$ be the weights realizing the maximum $\sum_S w_S = \overline{R}_\rho(x)$.
    By definition,
    \[
        \overline{R}_\rho(x) = \sum {w_S} \leq \sum_S{\rho(S) p^{-|S|} x_S} = R_\rho(x) \; .
    \]
    This proves the first item, and the second one follows by taking expectation on both sides of the inequality.

    For the third item, notice that if $x \leq y$, then for every set $x_S \leq y_S$.
    Hence, the feasible set of weights for $x$ is also feasible for $y$, as
    \[
        w_S \leq \rho(S) p^{-|S|} x_S \leq \rho(S) p^{-|S|} y_S \; ,
    \]
    and the claim follows.
    
    For the fourth item, let $\{w^*_S\}$ be the weights realizing the maximum $\sum_S w^*_S = \overline{R}_\rho(x^{(i \to 1)})$.
    Define:
    \[
        w'_S=
            \begin{cases}
            0, & i\in S, \\
            w^*_S, & i\notin S.
            \end{cases}
    \]

    Now, the set of weights $\{ w'_S\}$ is a feasible solution for $\overline{R}_\rho(x^{(i \to 0)})$.
    Hence,
    \[
        \overline{R}_\rho(x^{(i \to 0)}) \geq \sum_{S}{w'_S} = \sum_{S \notin \langle i \rangle}{w^*_S} \;,
    \]
    and then
    \begin{align*}
        \Delta_i \overline{R}_\rho (x^{(i)})
        &= \overline{R}_\rho (x^{(i \to 1)}) - \overline{R}_\rho (x^{(i \to 0)})  \\
        &\leq \sum_{S}{w^*_S} - \sum_{S \notin \langle i \rangle}{w^*_S} \\
        &= \sum_{S \in \langle i \rangle}{w^*_S} \\
        &\leq \Gamma \; ,
    \end{align*}
    where the last inequality follows from the cap constraint.
    In addition, $\overline{R}_\rho (x^{(i \to 1)}) \geq \overline{R}_\rho (x^{(i \to 0)})$ due to the monotonicity in $x$ (the previous item), and so $\Delta_i \overline{R}_\rho (x^{(i)}) \geq 0$.

    For the last item, observe that the weights $\{w_S\}$ are a feasible solution for $\overline{R}_{\rho'}(x)$, since for every $S$, $w_S \leq \rho(S)p^{-|S|}x_S \leq \rho'(S)p^{-|S|}x_S$.
    Therefore, $\overline{R}_{\rho'}(x) \geq \sum_S{w_S} = \overline{R}_\rho(x)$.
\end{proof}

\subsection{The CHT estimator is self-bounding} \label{section:self-bounding}
The main notion we use to prove the concentration inequalities for the CHT estimator is that of \emph{(strong) self-bounding functions} (cf. \cite[Section 6.11]{BLM13}).
\begin{definition} [strong $(a, b)$-self-bounding]
    A nonnegative function $f : \{0, 1\}^n \rightarrow [0, \infty)$ is called \emph{strongly $(a, b)$-self-bounding} for $a, b \geq 0$ if there exist functions  $f_i : \{0, 1\}^{n-1} \rightarrow [0, \infty)$ such that for all $i \in [n]$ and $x \in \{0, 1\}^n$:
    \[
        0 \leq f(x) - f_i \left( x^{(i)} \right) \leq 1\;,
    \]
    and
    \[
        \sum_{i=1}^n{\left( f(x) - f_i \left( x^{(i)}\right) \right)} \leq af(x) + b \; .
    \]
\end{definition}

\begin{claim} [CHT estimator is strongly self-bounding] \label{claim:cht-self-bounding}
    Let $\mu$ be a sub-distribution supported only on sets of size at most $q$.
    Let $\overline{R}_\mu$ be the $\Gamma$-Capped $p$-Horvitz--Thompson estimator for $\Gamma > 0$ and $p \in (0, 1]$.
    Then the function $f = \frac{\overline{R}_\mu}{\Gamma}$ is $(q, 0)$-strongly self-bounding.\footnote{When $\Gamma \leq 1$, $\overline{R}_\mu$ is also $(q, 0)$-strongly self-bounding.
    However, using normalization allows us to use stronger concentration bounds.}
\end{claim}

Before proving the claim, we make the following observation.
\begin{observation} \label{observation:derivative}
    For any nonnegative function  $f: \{0, 1\}^n \rightarrow [0, \infty)$ and any $x \in \{0, 1\}^n, i \in [n]$,
    \[
        x_i \Delta_i f(x^{(i)}) = f(x) - f(x^{(i \to 0)}) \; .
    \]
\end{observation}
\begin{proof}
    By definition,
    \[
         x_i \Delta_i f(x^{(i)}) = x_i \left( f(x^{(i \to 1)}) - f(x^{(i \to 0)}) \right) \;.
    \]
    
    If $x_i = 0$, then $x = x^{(i \to 0)}$ and $x_i \Delta_i f(x^{(i)}) = f(x) - f(x^{(i \to 0)}) = 0$.
    
    If $x_i = 1$, then $x = x^{(i \to 1)}$ and $x_i \Delta_i f(x^{(i)}) = f(x^{(i \to 1)}) - f(x^{(i \to 0)})$.
\end{proof}

The following two properties immediately prove \pref{claim:cht-self-bounding}.
\begin{claim} \label{claim:cht-bounds}
    Let $\mu$ be a sub-distribution supported only on sets of size at most $q$.
    Let $\overline{R}_\mu$ be the $\Gamma$-Capped $p$-Horvitz--Thompson estimator for $\Gamma > 0, p \in (0, 1]$.
    Then, for any $x \in \{0, 1\}^n, i \in [n]$,
    \begin{enumerate}
        \item $0 \leq \Delta_i \overline{R}_\mu(x^{(i)}) \leq \Gamma$.
        \item $\sum_{i=1}^n x_i \Delta_i \overline{R}_\mu(x^{(i)}) \leq q \overline{R}_\mu(x)$.
    \end{enumerate}
\end{claim}
We remark that the first item indicates that the derivatives of a $\Gamma$-capped HT estimator are indeed capped by $\Gamma$.

\begin{proof}
    First,  $\Delta_i \overline{R}_\mu(x^{(i)}) \geq 0$ by the monotonicity of $\overline{R}_\mu$ in $x$ (\pref{claim:cht-properties}).

    \paragraph{Bounding a single derivative. }
    To see that $\Delta_i \overline{R}_\mu(x^{(i)}) \leq \Gamma$, let $\{w^*_S\}$ be the weights realizing the maximum
    \[
        \sum_S w^*_S = \overline{R}_\mu(x^{(i \to 1)}) \; .
    \]
    We next claim that
    \[
        \sum_{S \notin \langle i \rangle}{w^*_S} \leq \overline{R}_\mu(x^{(i \to 0)}) \; .
    \]
    Define:
    \[
        w'_S=
            \begin{cases}
            0, & i\in S, \\
            w^*_S, & i\notin S.
            \end{cases}
    \]

    The set of weights $\{ w'_S\}$ is a feasible solution for $\overline{R}_\mu(x^{(i \to 0)})$:
    For every set $S$, we have $0 \leq w'_S \leq \mu(S) p^{-|S|}(x^{(i \to 0)})_S$, since if $i \in S$ then both the bound and $w'_S$ equal zero, and if $i \notin S$ then $w'_S = w^*_S$, which is bounded since $\{w^*_S\}$ is feasible.
    In addition, for every $j$, $\sum_{S \in \langle j \rangle} w'_S \leq \sum_{S \in \langle j \rangle} w^*_S \leq \Gamma$, so the cap constraints also hold. 

    Hence,
    \[
        \overline{R}_\mu(x^{(i \to 0)}) \geq \sum_{S}{w'_S} = \sum_{S \notin \langle i \rangle}{w^*_S} \;.
    \]

    And
    \begin{align*}
        \Delta_i \overline{R}_\mu (x^{(i)})
        &= \overline{R}_\mu (x^{(i \to 1)}) - \overline{R}_\mu (x^{(i \to 0)})  \\
        &\leq \sum_{S}{w^*_S} - \sum_{S \notin \langle i \rangle}{w^*_S} \\
        &= \sum_{S \in \langle i \rangle}{w^*_S} \\
        &\leq \Gamma \; .
    \end{align*}

    \paragraph{Bounding the derivatives sum. }
    The proof of the second item follows a similar structure.

    Let $\{w_S\}$ be the weights realizing the maximum
    \[
        \sum_S w_S = \overline{R}_\mu(x) \; ,
    \]
    and for every $i \in [n]$ define
    \[
        w^{(i)}_S=
            \begin{cases}
            0, & i\in S, \\
            w_S, & i\notin S.
            \end{cases}
    \]

    Now, for every $i \in [n]$ the set of weights $\{w^{(i)}_S\}$ is also a feasible solution for $\overline{R}_\mu(x^{(i \to 0)})$, and hence
    \[
        \overline{R}_\mu(x^{(i \to 0)}) \geq \sum_{S}{w^{(i)}_S} = \sum_{S \notin \langle i \rangle}{w_S} \;.    
    \]

    We get that
    \begin{align*}
        \sum_{i=1}^n x_i \Delta_i \overline{R}_\mu(x^{(i)})
        &= \sum_{i=1}^n{\left( \overline{R}_\mu(x) -\overline{R}_\mu(x^{(i \to 0)}) \right)} \\
        &\leq \sum_{i=1}^n \left( {\sum_S w_S -\sum_{S \notin \langle i \rangle}{w_S}} \right) \\
        &= \sum_{i=1}^n \sum_{S \in \langle i \rangle}w_S \\ 
    \end{align*}

   By changing the summation order, and since $\mu$ is supported on sets of size at most $q$,
   \[
    \sum_{i=1}^n x_i \Delta_i \overline{R}_\mu(x^{(i)}) \leq  \sum_{S}\sum_{i \in S} w_S = \sum_S |S| w_S \leq q \sum_Sw_S =  q \overline{R}_\mu(x) \; ,
   \]
    as required.
\end{proof}

\begin{proof} [Proof of \pref{claim:cht-self-bounding}]
    Let $x \in \{0, 1\}^n$, and define $f_i(x^{(i)}) = \frac{\overline{R}_\mu(x^{(i \to 0)})}{\Gamma}$.
    By the first item of \pref{claim:cht-bounds},
    \[
        \frac{\overline{R}_\mu(x)}{\Gamma} - \frac{\overline{R}_\mu(x^{(i \to 0)})}{\Gamma} =  \frac{x_i \Delta_i \overline{R}_\mu(x^{(i)})}{\Gamma} \leq 1 \; .
    \]
    Similarly, $\frac{x_i \Delta_i \overline{R}_\mu(x^{(i)})}{\Gamma} \geq 0$.

    By the second item of \pref{claim:cht-bounds},
    \[
        \sum_{i=1}^n \left( \frac{\overline{R}_\mu(x)}{\Gamma} - \frac{\overline{R}_\mu(x^{(i \to 0)})}{\Gamma} \right)
        =  \sum_{i=1}^n{\frac{x_i \Delta_i \overline{R}_\mu(x^{(i)})}{\Gamma}}
        \leq q f(x) \; .
    \]
\end{proof}

\subsection{Concentration of the CHT estimator} \label{section:concentration}
We now rely on the machinery of self-bounding functions to prove the concentration of the capped Horvitz--Thompson estimator around its mean.

\begin{theorem} [{\cite[Theorem~6.19]{BLM13}}] \label{thm:blm-upper-tail}
    Let $X =(X_1,\ldots , X_n)$ be a vector of independent random variables, each taking values in $\{0, 1\}$.
    Let $a > 0$, and let $f : \{0, 1\}^n \rightarrow [0, \infty)$ be a $(a, 0)$-strongly self-bounding function.\footnote{The theorem is stated in \cite{BLM13} under slightly more general assumptions, which are immediately satisfied by any $(a,0)$-strongly self-bounding function.}
    Let $Z = f (X)$.
    For all $t > 0$:
    \[
        \Pr[Z \geq \E[Z] + t]  \leq \exp \left( - \frac{t^2}{2a\E[Z] +at} \right) \; .
    \]
\end{theorem}

\begin{theorem} [{\cite[Special case of Theorem~6.20]{BLM13}}] \label{thm:blm-lower-tail}
    Let $X =(X_1,\ldots , X_n)$ be a vector of independent random variables, each taking
    values in $\{0, 1\}$.
    
    Let $a \geq 1/3$, and let $f : \{0, 1\}^n \rightarrow [0, \infty)$ be a $(a, 0)$-strongly self-bounding function.\footnote{Again, the theorem is stated under more general assumptions, which are satisfied by any $(a,0)$-strongly self-bounding function.}
    Let $Z = f (X)$.
    For $0 < t \leq \E[Z]$,
    \[
        \Pr[Z \leq \E[Z] - t]  \leq \exp \left( - \frac{t^2}{2a\E[Z]} \right) \; .
    \]
\end{theorem}

\begin{corollary} [Concentration of the CHT estimator] \label{corollary:cht-concentration}
    Let $X =(X_1,\ldots , X_n)$ be a vector of independent Bernoulli random variables with probability $p$.

    Let $\mu$ be a sub-distribution supported only on sets of size at most $q$, let $\theta \in (0, 1)$, and let $\gamma > 0$.
    Let $\overline{R}_\mu$ be the $\Gamma$-Capped $p$-Horvitz--Thompson estimator for $\mu$, and denote $Z = \overline{R}_\mu(X)$.

     If $\E[Z] \leq \theta - \gamma$, then
     \[
        \Pr[Z \geq \theta] \leq \exp\left(-\frac{\gamma^2}{2q\Gamma}\right) \; .
     \]

    If $\E[Z] \geq \theta + \gamma$, then
    \[
        \Pr[Z \leq \theta] \leq \exp\left(-\frac{\gamma^2}{2q\Gamma}\right) \; .
    \] 
\end{corollary}

\begin{proof}
    By \pref{claim:cht-self-bounding},  $f = \frac{\overline{R}_\mu}{\Gamma}$ is a $(q,0)$-strongly self-bounding function.
    Let
    \[
        Y=\frac{Z}{\Gamma}=f(X),
    \]
    and
    \[
        t = \frac{\gamma}{\Gamma} \;.
    \]
    
    \paragraph{Upper tail. }
    Suppose $\E[Z] \leq \theta - \gamma$, and thus $\frac{\theta}{\Gamma} \geq \E[Y] + \frac{\gamma}{\Gamma}$.

    Applying \pref{thm:blm-upper-tail} with $t > 0$ and $a = q$, we get
    \begin{align*}
        \Pr[Z \geq \theta]
        &=  \Pr[Y \geq \frac{\theta}{\Gamma}] \\ 
        &\leq \Pr[Y \geq \E[Y] + \frac{\gamma}{\Gamma}] \\ 
        &\leq \exp \left( -\frac{(\gamma/\Gamma)^2}{2q\E[Y] +q\gamma/\Gamma}  \right) \\ 
        &= \exp \left( -\frac{\gamma^2}{q\Gamma(2 \Gamma\E[Y]+\gamma)}  \right) \\ 
        &\leq  \exp \left( -\frac{\gamma^2}{2q\Gamma} \right) \; ,
    \end{align*}
    where the last inequality follows from $2\Gamma \E[Y] + \gamma =2\E[Z] + \gamma \leq 2(\theta - \gamma) + \gamma < 2$, as $\theta < 1$ and $\gamma > 0$.
    
    \paragraph{Lower tail. }
    Suppose $\E[Z] \geq \theta + \gamma$, and thus $\E[Y] \geq \frac{\theta + \gamma}{\Gamma} \geq \frac{\gamma}{\Gamma} = t$.
    We have $q \geq 1 > 1/3$, and $t \leq  \E[Y]$, so the hypothesis of \pref{thm:blm-lower-tail} is satisfied with $t$ and $a = q$.
    We get
    \begin{align*}
           \Pr[Z \leq \theta]
           &=  \Pr[Y \leq \frac{\theta}{\Gamma}] \\ 
           &\leq \Pr[Y \leq \E[Y] -\frac{\gamma}{\Gamma}] \\ 
           &\leq \exp \left( -\frac{(\gamma/\Gamma)^2}{2q\E[Y]} \right) \\ 
           &\leq \exp \left( -\frac{\gamma^2}{2q\Gamma} \right) \; ,
    \end{align*}
    where the last inequality follows from $\E[Y] = \E[Z] / \Gamma\leq 1/\Gamma$.
\end{proof}

\subsection{Spreadness implies $\E[R]$ is almost $\E[\overline{R}]$} \label{section:expectation}
    We have shown that the \emph{capped} HT estimator concentrates around its mean.
    It is left to show that so long as the underlying sub-distribution is \emph{spread}, the expectation of the capped HT estimator is not much smaller than that of the (non-capped) HT estimator, which is the quantity we would like to estimate.

We need the following elementary inequality.
\begin{lemma} \label{lemma:var}
    Let $X$ be a random variable with finite variance, and let $h > 0$.
    Suppose that $\E[X] \leq h$.
    Then,
    \[
        \E[(X-2h)_+] \leq \frac{\Var(X)}{4h} \; .
    \]
\end{lemma}
\begin{proof}
    Let $h^* := \E[X]$.
    We first argue that for every $x$,
    \[
        (x-2h)_+ =\max(x-2h,0) \leq \frac{(x-h^*)^2}{4h} \; .
    \]
    If $x \leq 2h$, then the left hand side of the inequality is zero, and the inequality holds.
    Otherwise, denote $x = h(1+t)$ for $t > 1$.
    We now have $x-2h = h(t - 1)$, and $x-h^* \geq x-h =ht$.
    
    Therefore, it is sufficient to show that:
    \[
        h(t-1)  \leq \frac{(ht)^2}{4h} \iff 4t-4  \leq t^2 \iff t^2 -4t + 4 \geq 0 \; .
    \]
    This is equivalent to $(t-2)^2 \geq 0$, which is true.

    Next, we take expectation on both sides of the inequality, and get
    \[
        \E[(X-2h)_+] \leq \frac{\E[(X-h^*)^2]}{4h} = \frac{\Var(X)}{4h} \; .
    \]
\end{proof}

\begin{lemma} \label{lemma:cht-expectation}
    Let $\mu$ be an $(m, r)$-spread sub-distribution over $\mathcal{P}([n])$ supported on sets of size at most $q$. 
    Let $p \in (0, 1]$, let $R_\mu$ be the $p$-Horvitz--Thompson estimator for $\mu$, and let $\overline{R}_\mu$ be the $\Gamma$-Capped $p$-Horvitz--Thompson estimator for $\mu$ with $\Gamma=2m/pr$.
    Further assume that $pr \geq 2(q-1)$.
    
    Let $X =(X_1,\ldots , X_n)$ be a vector of independent Bernoulli random variables with probability $p$.
    Then,
    \[
        \E[R_\mu(X)] - \E[\overline{R}_\mu(X)]= O \left( \frac{q^2}{pr}\right) \; .
    \]
\end{lemma}

\begin{proof}
    For simplicity of notation, we remove the $\mu$ subscripts.

    We analyze the random variables $\Delta_i R(X^{(i)})$, and relate them to the expectation gap.
    Specifically, we prove the following three claims:
    \begin{claim} \label{claim:gap-bound}
        For every $x \in \{0, 1\}^n$,
         \[
            R(x) - \overline R(x) \leq \sum_{i \in [n]} \left(x_i \Delta_i R(x^{(i)}) - \Gamma \right)_+ .
        \]        
    \end{claim}
    
    \begin{claim} \label{claim:ht-mean}
        For every $i \in [n]$,
        \[
            \E[\Delta_i R(X^{(i)})] \leq \frac{m}{pr} = \frac{\Gamma}{2} \; .
        \]
    \end{claim}

    \begin{claim} \label{claim:ht-var}
        For every $i \in [n]$,
         \[
            \Var(\Delta_i R(X^{(i)})) \leq \frac{2m(q-1)} {p^3r^2} \cdot \mu(\langle i\rangle) \; .
        \]
    \end{claim}
    Before proving the claims, let us show that combined with \pref{lemma:var} they imply the Lemma.

    First, taking expectation on the inequality of the first claim, we get
    \[
        \E[R(X)] - \E[\overline{R}(X)] \leq \sum_{i \in [n]} \E\left[ \left(X_i \Delta_i R(X^{(i)}) - \Gamma \right)_+ \right] \; .    
    \]

    Now, by \pref{lemma:var}, and as $\E[\Delta_i R(X^{(i)})] \leq \frac{\Gamma}{2}$, we get
    \[
         \E[(\Delta_i R(X^{(i)})-\Gamma)_+] \leq \frac{\Var(\Delta_i R(X^{(i)}))}{4 \cdot \Gamma/2} = \frac{p r}{4m} \Var (\Delta_i R(X^{(i)}))
    \]
    Now, since $X_i$ is independent of $\Delta_i R(X^{(i)})$, we have
    \[
        \E[(X_i \Delta_i R(X^{(i)}) - \Gamma)_+] = p \E[(\Delta_i R(X^{(i)})-\Gamma)_+] \leq \frac{p^2 r}{4m} \Var (\Delta_i R(X^{(i)})) \; .
    \]

    Summing over all $i \in [n]$, and using the bound on $ \Var (\Delta_i R(X^{(i)}))$, we deduce:
    \begin{align*}
        \E[R(X)] - \E[\overline{R}(X)]
        &\leq \sum_{i \in [n]} \E \left[ \left(X_i \Delta_i R(X^{(i)}) - \Gamma \right)_+ \right] \\
        &\leq \sum_{i \in [n]} \frac{p^2 r}{4m} \Var (\Delta_i R(X^{(i)})) \\
        &\leq  \sum_{i \in [n]} \frac{p^2 r}{4m} \cdot \frac{2m(q-1)} {p^3r^2} \cdot \mu(\langle i\rangle) \\
        &= \frac{q-1}{2pr}  \sum_{i \in [n]}  \mu(\langle i\rangle) \; .
    \end{align*}

    Opening the definition of $\mu(\langle i \rangle)$, and since $\mu$ is supported on sets of size at most $q$, we get:
    \[
        \sum_{i \in [n]} \mu(\langle i \rangle) = \sum_{i \in [n]} \sum_{S \in \langle i \rangle} \mu(S) = \sum_S{|S| \mu (S)} \leq q \sum_S \mu(S) \leq q \; .
    \]
    We conclude that 
    \[
          \E[R(X)] - \E[\overline{R}(X)] \leq \frac{q-1}{2pr} \cdot q  = O\left(\frac{q^2}{pr} \right) \; .
    \]

    We now return to proving the stated claims.
    \begin{proof} [Proof of \pref{claim:gap-bound}]
        To bound $R(x) - \overline{R}(x)$, we construct a set of weights $\{w_S\}$, such that:
    \begin{enumerate}
        \item For every $S$, $0 \leq w_S \leq  \mu(S) p^{-|S|} x_S$.
        \item For every $i \in [n]$, $\sum_{S \in \langle i \rangle}w_S \leq \Gamma$.
        \item $R(x) - \sum_S{w_S} \leq \sum_{i \in [n]} \left(x_i \Delta_i R(x^{(i)}) - \Gamma \right)_+$.
    \end{enumerate}
    
    We construct the weights by ``trimming'' the weights of the Horvitz--Thompson estimator $R(x)$.
    Start with $w_S = \mu(S) p^{-|S|} x_S$.
    Now, iterate over each $i \in [n]$.
    If $\sum_{S \in \langle i \rangle}w_S \leq \Gamma$ is already satisfied, continue to the next index.
    Otherwise, reduce the weights $\{w_S\}_{S \in \langle i \rangle}$ by a total amount of
    \[
        \sum_{S \in \langle i \rangle}w_S - \Gamma \; ,
    \]
    distributing the reduction arbitrarily among these weights while keeping them non-negative.

    The inequality $w_S \leq  \mu(S) p^{-|S|} x_S$ is satisfied for all $S$, since we only reduce the weights.
    Similarly, $\sum_{S \in \langle i \rangle}w_S \leq \Gamma$ is satisfied for every $i$, since we only reduce weights - reductions while processing some index can only ``help'' other indices.

    Now, throughout the construction, we have $\sum_{S \in \langle i \rangle} w_S \leq  \sum_{S \in \langle i \rangle}\mu(S) p^{-|S|} x_S = x_i \Delta_i R(x^{(i)})$.
    While processing index $i$, we reduce $\sum_S w_S$ by $\left(\sum_{S \in \langle i \rangle} {w_S} - \Gamma \right)_+ \leq \left(x_i \Delta_i R(x^{(i)}) - \Gamma \right)_+$, hence (3) is also satisfied.

    From (1) and (2) we conclude that $\{ w_S \}$ are a feasible solution to the maximization problem defining $\overline{R}(x)$, and hence $\overline{R}(x) \geq \sum_S w_S$.
    This inequality combined with (3) completes the argument.
    \end{proof}

    \begin{proof} [Proof of \pref{claim:ht-mean}]
        By a direct computation,
        \begin{align*}           
            \E[ \Delta_i R(X^{(i)})] &= \E[\sum_{S \in \langle i \rangle}{\mu(S)\cdot p^{-|S|} X_{S \setminus \{ i \} }}] \\
            &= \sum_{S \in \langle i \rangle}{\mu(S)\cdot p^{-|S|} \E[X_{S \setminus \{ i \} }]} \\ 
            &= \frac{1}{p} \sum_{S \in \langle i \rangle}{\mu(S)} \\ 
            &= \frac{\mu(\langle i \rangle)}{p} \; .  
        \end{align*}

    Since $\mu$ is $(m, r)$-spread, we have $\mu(\langle i \rangle) \leq \frac{m}{r}$, and hence 
    \[
        \E[\Delta_i R(X^{(i)})] \leq \frac{m}{pr} = \frac{\Gamma}{2} \; .
    \]
    \end{proof}
    
    \begin{proof} [Proof of \pref{claim:ht-var}]
    We next bound the variance of $\Delta_i R(X^{(i)})$.
    First, 
    \begin{align*}
        \E[\left( \Delta_i R(X^{(i)}) \right)^2] &= \E[\sum_{S,T \in \langle i \rangle}{\mu(S)\mu(T) p^{-|S|-|T|}X_{(S \cup T) \setminus \{i\}}}] \\
        &= \sum_{S,T \in \langle i \rangle}{\mu(S)\mu(T) p^{-|S|-|T|}p^{|(S \cup T) \setminus \{i\}|}} \\
        &= \frac{1}{p^2} \sum_{S,T \in \langle i \rangle}{\mu(S)\mu(T) p^{-|(S \cap T) \setminus \{i\}|}} \; .
    \end{align*}

    And since $\E[ \Delta_i R(X^{(i)})]^2 = \frac{\mu(\langle i \rangle)^2}{p^2}$ we get
    \[
        \Var(\Delta_i R(X^{(i)})) =\frac{1}{p^2} \sum_{S,T \in \langle i \rangle}{\mu(S)\mu(T) (p^{-|(S \cap T) \setminus \{i\}|} -1)} \; .
    \]

    We now iterate over all the pairs of sets $S, T \in \langle i \rangle$ as follows:
    first, we iterate over every $S  \in \langle i \rangle$.
    Second, over every intersection size $t \in [q]$.
    Next, over the possible intersection sets $R$ (that is, subsets of $S$) of size $t$ that contain $i$.
    Finally, over every $T \in \langle i \rangle$ such that $S \cap T = R$.
    The last iteration is where we leverage the spreadness, as the total mass of sets $T$ that include a subset $R$ is bounded.

    We get that:
    \begin{align*}
        \sum_{S,T \in \langle i \rangle}{\mu(S)\mu(T) (p^{-|(S \cap T) \setminus \{i\}|} -1)} &= 
         \sum_{S\in \langle i \rangle} \sum_{t=1}^q \sum_{\substack{R\in {S\choose t}\\i\in R}} \sum_{T: S\cap T = R} {\mu(S)\mu(T) (p^{-|(S \cap T) \setminus \{i\}|} -1)} \\
         &= \sum_{S\in \langle i \rangle} \mu(S)\sum_{t=1}^q \sum_{\substack{R\in {S\choose t}\\i\in R}} \sum_{T: S\cap T = R} {\mu(T) (p^{-t+1} -1)} \\
         &= \sum_{S\in \langle i \rangle} \mu(S)\sum_{t=1}^q (p^{-t+1} -1) \sum_{\substack{R\in {S\choose t}\\i\in R}} \sum_{T: S\cap T = R} {\mu(T)} \; .
    \end{align*}

    Now, $S \cap T = R \implies T \supseteq R$, and so
    \[
        \sum_{T: S\cap T = R} \mu(T) \leq \sum_{T \supseteq R} {\mu(T)} = \mu(\langle R \rangle) \; ,
    \]
    where the last equality is by definition.

    We now use the spreadness assumption: if $|R| = t$, then $\mu(\langle R \rangle) \leq \frac{m}{r^t}$.
    Hence, and as $|S| \leq q$,
    \[
        \sum_{\substack{R\in {S \choose t}\\i\in R}} \mu(\langle R \rangle)
        \leq \sum_{\substack{R\in {S \choose t}\\i\in R}} \frac{m}{r^t} = \frac{m}{r^t} {|S| - 1 \choose t-1} \leq \frac{m}{r^t} (|S| - 1)^{t-1} \leq \frac{m}{r^t} (q - 1)^{t-1}\; .
    \]

    Next, we calculate the remaining sum over $t$.
    
    We can skip the first term, with $t = 1$, since $p^{-t+1} -1 = 1-1 = 0$.
    In addition, for $t > 1$ we have $p^{-t+1} - 1 \leq p^{-t+1}$.
    Combining this with the previous calculations, we deduce
    \[
        \sum_{t=1}^q{(p^{-t+1}-1) \frac{m}{r^t} (q-1)^{t-1}}
        \leq \frac{m}{r} \sum_{t=2}^q {\left(\frac{q-1}{pr} \right)^{t-1}}
        \leq \frac{m}{r} \sum_{t=2}^\infty {\left(\frac{q-1}{pr} \right)^{t-1}} \; .
    \]

    By the formula for geometric sums (\pref{fact:geometric-sum}),
    \[
        \sum_{t=2}^\infty {\left(\frac{q-1}{pr} \right)^{t-1}} 
        = \sum_{t=1}^\infty {\left(\frac{q-1}{pr} \right)^{t}}
        = \frac{q-1}{pr-q+1}
        \leq \frac{2(q-1)}{pr} \; ,
    \]
    where the last inequality is by the assumption $pr \geq 2(q-1)$.

    Finishing the calculations, and since $\sum_{S \in \langle i \rangle} \mu(S) = \mu(\langle i \rangle)$, we get:
    \begin{align*}
        \Var(\Delta_i R(X^{(i)}))
        &= \frac{1}{p^2} \sum_{S,T \in \langle i \rangle}{\mu(S)\mu(T) (p^{-|(S \cap T) \setminus \{i\}|} -1)} \\
        &\leq \frac{1}{p^2} \sum_{S \in \langle i \rangle} \mu(S) \cdot \frac{m}{r} \cdot\frac{2(q-1)}{pr} \\
        &= \frac{1}{p^2} \cdot \mu(\langle i \rangle) \cdot \frac{2m(q-1)}{pr^2} \\
        &= \frac{2m(q-1)}{p^3r^2}\mu(\langle i \rangle) \; .
    \end{align*}
    \end{proof}
\end{proof}

\section{Proof of the Lower Bound}
In this section, we prove our main theorem.
    \begin{theorem} \label{thm:main}
        Suppose $C :\{0,1\}^k\rightarrow \{0, 1\}^n$ is a $(q, \delta, \alpha, \beta)$-RLDC where $\delta, \alpha, \beta$ are constants.
        Suppose that $\alpha + \beta > 1$, and denote $\lambda = \frac{\alpha}{\alpha + \beta - 1}$.
        Then, for every constant $\varepsilon > 0$,
        \[
            n=\Omega_{\varepsilon, \alpha, \beta} \left(q^{-2} \left(\frac{k}{\log k}\right)^{1+\frac{1}{(\lambda + \varepsilon)q}} \right) \; .
        \]
    \end{theorem}
    \pref{thm:main-intro} is a direct corollary of \pref{thm:main}, with $\alpha=\beta=2/3$.

    The requirement  $\alpha + \beta >1$ is necessary for a superlinear lower bound.
    For any $\alpha$ and $\beta$ satisfying $\alpha + \beta \leq 1$, there exists a trivial 1-query RLDC, with $n=k$ and an arbitrarily large decoding radius.
    Specifically, consider the identity code $C(x) = x$ and the relaxed decoder that operates as follows:
    to decode $x_i$ from an input $y$, it outputs $y_i$ with probability $\alpha$ and $\bot$ with probability $1-\alpha$.
    Completeness holds with probability $\alpha$, and soundness holds, for any $y$, with probability at least $1-\alpha \geq \beta$.\footnote{For an RLDC with non-trivial relative distance, one can instead use any systematic constant-rate code with constant relative distance.}

The proof of \pref{thm:main} has two parts.
First, we define the notion of a \emph{sample-based decoder} of a code, and prove that having such a decoder implies a lower bound on the blocklength of the code.
Second, we show how to construct a sample-based decoder for any RLDC.

\subsection{RLDC Lower Bounds from Sample-Based Decoders}
A main idea in the proof is that of sample-based decoders, which we define next. 

\begin{definition} [Sample-Based Decoder] \label{def:global-decoder}
    Let $C :\{0,1\}^k\rightarrow \{0, 1\}^n$.
    A \emph{sample-based decoder} with sampling probability $p \in (0, 1]$ and error probability $\varepsilon \in (0, 1/2)$ for $C$ is a randomized algorithm  that operates as follows:
    
    Given a codeword $c = C(x)$, the algorithm samples a set $W \subseteq [n]$ by including each index independently with probability $p$, reads $c|_W$, and outputs $\hat{x} \in \{0,1\}^k$, which is a deterministic function of $W$ and $c|_W$.
    A sample-based decoder must satisfy the following requirement:

    For every $x \in \{0,1\}^k$ and every $i \in [k]$,
        \[
            \Pr[\hat{x}_i \neq x_i] \leq \varepsilon \; ,
        \]
    where the probability is over the choice of $W$.
\end{definition}

Having a sample-based decoder implies a lower bound on the block-length of the code.
\begin{lemma} \label{lemma:decoder-to-lb}
    Let $C : \{0,1\}^k \rightarrow \{0,1\}^n$.
    Suppose that $C$ has a sample-based decoder with sampling probability $p$ and error probability $\varepsilon$.
    Then
    \[
        pn \geq (1-h_2 (\varepsilon ))k \; ,
    \]
    where $h_2$ denotes the binary entropy function.
    In particular, if $\varepsilon < 1/2$ is constant, then $pn=\Omega(k)$.
\end{lemma}

\begin{proof}
    Let $X$ be uniform over $\{0,1\}^k$, let $W$ be the sampled set, and denote
    \[
        Y=(W,C(X)|_W) \; .
    \]
    For every $i\in[k]$, the decoder predicts $X_i$ from $Y$ with error at most $\varepsilon$.
    Hence, by Fano's inequality (\pref{fact:fano}),

    \[
        H(X_i \mid Y) \leq h_2(\varepsilon) \; .
    \]
    By subadditivity of conditional entropy,
    \[
        H(X \mid Y) \leq \sum_i H(X_i \mid Y) \leq  kh_2(\varepsilon) \;.  
    \]

    On the other hand, conditioned on $W=w$, the decoder observes $C(X)|_w$.
    Since $C(X)|_w$ is a string of length $|w|$, its entropy satisfies $H(C(X)|_w \mid W = w) \leq |w|$.
    Hence,
    \begin{align*}
        H(X \mid W=w, C(X)|_w) &= H(X \mid W = w) - H(C(X)|_w \mid W=w) \\
        &\geq H(X\mid W=w)-|w| \; ,
    \end{align*}
    where the equality follows from the chain rule and the fact that $C(X)|_w$ is a deterministic function of $X$ for fixed $w$ (\pref{fact:chain}).
    
    Since $W$ is independent of $X$, we have $H(X\mid W=w)=H(X)=k$.
    Averaging over $W$, we get
    \[
        H(X \mid Y) \geq H(X) - \E[|W|]  = k-pn \; ,
    \]
    where $H(X) = k$ since $X$ is chosen uniformly over $\{0, 1\}^k$.

    Combining the two bounds gives
    \[
        k -pn \leq k h_2(\varepsilon) \;.
    \]
    Therefore,
    \[
        pn \geq (1-h_2(\varepsilon))k \;.
    \]
\end{proof}

\subsection{Sample-Based Decoders from RLDCs}
The following Lemma states that if a code has a relaxed decoder, then it also has a sample-based decoder.
\begin{lemma} \label{lemma:global-decoder}
    Let $C :\{0,1\}^k\rightarrow \{0, 1\}^n$ be a $(q, \delta, \alpha, \beta)$-RLDC where $\delta, \alpha, \beta$ are constants satisfying $\alpha + \beta > 1$.
    Let $\varepsilon > 0$ be a fixed constant, let $d= \lfloor (\frac{\alpha}{\alpha + \beta - 1} + \varepsilon)q\rfloor + 1$ and let $p=\min(1, D \frac{q^2\log n}{n^{1/d}})$ for a sufficiently large constant $D = D(\varepsilon, \alpha, \beta)$.

    Then, for sufficiently large $n$, $C$ has a sample-based decoder with sampling probability $p$ and error probability at most $1/3$.
\end{lemma}

    \begin{remark} \label{remark:wlog-p}
    Notice that if $p = 1$ (that is, $n^{1/d} \leq D q^2 \log n$), then the sample-based decoder reads \emph{all} of the $n$-bit codeword, and the lemma is trivial.
    Hence, without loss of generality, we assume $p<1$ throughout.
    \end{remark}

    In the remainder of this subsection we prove \pref{lemma:global-decoder}.

\subsubsection{Extracting Spreadness}
Before describing the operation of the sample-based decoder, let us set up the main tool on which we are based, the spreadness extraction lemma.

The following is a straightforward adaptation of \pref{lemma:large_spread_daisy} to sub-distributions.
\begin{corollary} (Spreadness Extraction for Sub-distributions) \label{corollary:large_spread_daisy2}
    Let $\mu$ be a sub-distribution over $\mathcal{P}_{\leq q}(U)$, where $|U| = n$.
    Let $d > q$ be any integer.
    Suppose that $\|\mu\|_1 > 0$. 
    Then, there exists $j \in [d]$, a family of sets $\D \subseteq \mathrm{supp}(\mu)$, and a set of elements $H \subseteq U$ such that the sub-distribution $(\mu|_\D)^{\circ H}$ is $(m, n^{1/d})$-spread, where:
    \begin{align*}        
        m &= q \frac{n^{j/d}}{n}, &
        |H| &= \left\lfloor \frac{n}{n^{j/d}} \right \rfloor = \left\lfloor \frac{q}{m} \right \rfloor, &
        \mu(\D) &\geq \left( 1 - \frac{q}{d} \right)\|\mu\|_1 \; ,
    \end{align*}
    where $\mu|_\D$ is the restriction of $\mu$ to $\D$. (It is the sub-distribution that gives weight $\mu(S)$ for every $S \in \D$, and weight zero for every other set.
    That is, $\mu|_\D = \mu(\D)\mu_\D$).
\end{corollary} 

\begin{proof}
    Define $\nu = \frac{\mu}{\|\mu\|_1}$.
    This is a distribution that satisfies the conditions of \pref{lemma:large_spread_daisy}.
    Hence, there exists $j \in [d]$, $\D \subseteq \supp(\mu)$ and $H' \subseteq U$ such that $(\nu_\D)^{\circ H'}$ is $(m', n^{1/d})$-spread, where:
    \begin{align*}        
        m' &= \frac{q}{\nu(\D)} \cdot \frac{n^{j/d}}{n}, &
        |H'| & \leq \frac{n}{n^{j/d}}, &
        \nu(\D) &\geq 1 - \frac{q}{d} \; .
    \end{align*}

    \paragraph{Padding. }
    First, we add to $H'$ exactly $\left\lfloor \frac{n}{n^{j/d}} \right \rfloor -|H'|$ arbitrary elements from $U \setminus H'$, and let the resulting set be $H$.
    The distribution $(\nu_\D)^{\circ H}$ is also $(m', n^{1/d})$-spread, since removing universe elements only relaxes the spreadness condition.

    \paragraph{Total mass. }
    Now, $\mu = \|\mu\|_1\nu$, and hence $\mu(\D) = \|\mu\|_1 \nu(\D) \geq \left( 1 - \frac{q}{d} \right)\|\mu\|_1$.

    \paragraph{Spreadness}
    First, notice that $\nu_\D = \mu_\D$, since for every $S \in \D$
    \[ 
        \nu_\D(S)=\frac{\nu(S)}{\nu(\D)} = \frac{\mu(S)/\|\mu\|_1}{\mu(\D)/\|\mu\|_1} = \frac{\mu(S)}{\mu(\D)} = \mu_\D(S) \; .
    \]
    In addition, by definition $\mu(\D) = \|\mu\|_1 \nu(\D)$.
    Therefore, we have
    \[
        \mu|_{\D} = \mu(\D)\nu_{\D} = \|\mu\|_1\nu(\D)\nu_{\D} \; .
    \]
    
    Hence, for every non-empty $T\subseteq U\setminus H$,
    \[
    \begin{aligned}
        ((\mu|_{\D})^{\circ H})(\langle T\rangle) &= \|\mu\|_1\nu(\D) ((\nu_{\D})^{\circ H})(\langle T\rangle) \\
        &\leq \|\mu\|_1\nu(\D) \frac{m'}{n^{|T|/d}} \\
        &= \|\mu\|_1 \frac{q n^{j/d}/n}{n^{|T|/d}} \\
        &\leq \frac{q n^{j/d}/n}{n^{|T|/d}} \; .
    \end{aligned}
    \]
    Thus, $(\mu|_\D)^{\circ H}$ is $\left( q \frac{n^{j/d}}{n}, n^{1/d}\right)$-spread.
\end{proof}

\subsubsection{Constructing the Sample-Based Decoder}
We next describe the operation of the stated sample-based decoder of \pref{lemma:global-decoder}.
First, we introduce some definitions.
\paragraph{Definitions. }
Let $T$ be a decision tree, and $y \in \{0, 1\}^n$.
\begin{enumerate}
    \item We denote by $T(y) \in \{0, 1, \bot\}$ the output of the decision tree when $y$ is its input.
    \item We denote by $Q_T(y) \subseteq [n]$ the set of indices queried on the active path of $T$ on $y$.
    \item Let $A \subseteq [n]$ and $a \in \{0, 1\}^A$.
    
    We denote by $y_{A \xleftarrow{} a} \in \{0, 1 \}^n$ the vector defined by $y_{A \xleftarrow{} a}(i) = \begin{cases} 
            a(i) & \text{if } i \in A\\ 
            y_i & \text{if } i \notin A
    \end{cases}$.
    \item We denote the indicator vector of a set $A \subseteq [n]$ by $\mathbf{1}_A \in \{0, 1\}^n$.
\end{enumerate}

\begin{construction} \label{construction:global-decoder}
    Recall that $d = \lfloor(\frac{\alpha}{\alpha + \beta - 1} + \varepsilon)q \rfloor + 1$.
    Set $r=n^{1/d}$, and let $M = \{q n^{\frac{j}{d} - 1} \mid j \in [d]\}$.\footnote{Notice that $M$ corresponds to the possible values $m$ can get in \pref{corollary:large_spread_daisy2}.}
    Note that with these definitions, we have $p \leq D \frac{q^2 \log n}{r}$, and $pr = O(q^2 \log n)$.

    \paragraph{Setting the threshold. }
    Let us denote
    \begin{align*}
        \Lambda &= \alpha+\beta - 1, &
        \lambda &= \frac{\alpha}{\Lambda}, &
        \eta &= \frac{\varepsilon \Lambda}{\lambda + \varepsilon} \;.
    \end{align*}

    Since 
    \[
        d = \lfloor(\lambda + \varepsilon)q \rfloor + 1 > (\lambda + \varepsilon)q \; ,
    \]
    the gap between $\alpha(1-q/d)$ and $1 - \beta$ is greater than $\eta$, a constant which depends only on $\varepsilon, \alpha, \beta$:
    \[
        \alpha(1-q/d) - (1 -\beta) > \frac{\varepsilon \Lambda}{\lambda + \varepsilon} > 0 \;.
    \]

    We choose the threshold to be the midpoint between these endpoints
    \[
        \theta = \frac{1-\beta + \alpha(1-q/d)}{2} \;.
    \]
    Note that $\alpha(1-q/d) > \theta + \eta/2$ and $1-\beta < \theta - \eta/2$.

    \paragraph{The sample-based decoder. }
    The sample-based decoder operates in two phases.
    First, as it must by its definition, it samples $W \sim \mathrm{Bin}([n], p)$, and queries its input $c$ at indices $W$.
    
    In the second phase, for every index $i \in [k]$, it computes deterministically from $(W, c|_W)$ a value $\hat{x}_i \in \{0, 1\}$.
    To describe this phase, and for the rest of the proof, let us fix $i \in [k]$ and let $\tau:=\tau_i$ be the distribution of trees that corresponds to the operation of the relaxed decoder for decoding index $i$.
    We formally describe the computation of $\hat{x}_i$ from $(W, c|_W)$ in \pref{algo:global_decoder}.

    \begin{algorithm}
    \caption{Sample-based decoder for the RLDC $C$ for a fixed index $i \in [k]$ \\
    Input: A pair $(W, c|_W)$ where $W \subseteq [n]$, $c=C(x)$, and $c|_W$ is the restriction of $c$ to $W$. \\
    Output: $b \in \{0, 1\}$}
    \label{algo:global_decoder}
        \begin{enumerate}
            \item For every $m \in M$, every $H \subseteq [n]$ such that $|H| = \left \lfloor \frac{q}{m} \right \rfloor$, and every $\kappa \in \{0, 1\}^H$:
            \begin{enumerate}
                \item Let $y = c_{H \xleftarrow{} \kappa}$.
                \item For every $b \in \{0, 1\}$:
                    \begin{enumerate}
                        \item Define the sub-distribution $\rho_{H, \kappa, b}: \P{([n] \setminus H)} \rightarrow [0, 1]$ by 
                            \[ 
                                \rho_{H, \kappa, b}(S) = \Pr_{T \sim \tau}[T(y) = b \ \land \ Q_T(y) \setminus H=S] \; .
                            \]
                        \item Let $\overline{R}_{m, H, \kappa, b}$ be the $\frac{2m}{pr}$-Capped $p$-Horvitz--Thompson estimator for $\rho_{H, \kappa, b}$.
                        \item Denote $X = \mathbf{1}_W$.
                        \item If $\overline{R}_{m, H, \kappa,b}(X) \geq \theta$, halt and output $b$.
                    \end{enumerate}                    
            \end{enumerate}
            \item If the algorithm did not halt, output $0$.
        \end{enumerate}        
    \end{algorithm}
    \end{construction}

    Notice that $\overline{R}_{m, H, \kappa,b}(X)$, as defined in the algorithm, depends on $c_{H \xleftarrow{} \kappa}$, which is unknown to the algorithm.
    Hence, before we even prove the correctness of the algorithm, we need to show that it can be executed.
    \begin{claim} [CHT estimator is determined by the sampled view]
        For every $m, H, \kappa, b, W$, the value $\overline{R}_{m, H, \kappa,b}(X)$ is determined by $c|_W$.
        That is, the values of $c$ at indices not in $W$ do not affect the CHT estimator.
    \end{claim}
\begin{proof}
    This follows from the definition of the capped Horvitz--Thompson estimator.

    For every set $S\not\subseteq W$, we have $X_S=0$, and therefore the constraints defining the capped Horvitz--Thompson estimator imply $w_S=0$.
    Hence, to compute $\overline{R}_{m, H, \kappa,b}(X)$, it suffices to compute $\rho_{H, \kappa, b}(S)$ for sets $S \subseteq W$.

    Fix such a set $S\subseteq W$.
    To determine whether a decision tree $T$ contributes to $\rho_{H,\kappa,b}(S)$, we simulate $T$ on $y=c_{H\xleftarrow{}\kappa}$.
    Queries in $H$ are answered using $\kappa$, and queries in $W \setminus H$ are answered using $c|_W$.
    If during the simulation $T$ queries an index outside $H \cup W$, then necessarily $ Q_T(y)\setminus H \neq S$, and $T$ does not contribute to $\rho_{H,\kappa,b}(S)$.
    Otherwise, the execution of $T$ can be determined from $\kappa$ and $c|_W$, and the algorithm determines whether $T(y)=b$ and $Q_T(y)\setminus H=S$.
\end{proof}

\subsubsection{Analysis of the Sample-Based Decoder}
The following two claims show the correctness of \pref{construction:global-decoder}, and prove \pref{lemma:global-decoder}.

Recall that $b^* = x_i$ is the bit the sample-based decoder needs to output.
\begin{claim} [Soundness] \label{claim:soundness}
    With probability at least $5/6$, for every $m \in M, H \in {[n] \choose \lfloor q/m \rfloor}$, $ \kappa \in \{0, 1\}^H$ we have $\overline{R}_{m, H, \kappa, (1-b^*)}(X) < \theta$.
\end{claim}

\begin{claim} [Completeness] \label{claim:completeness}
    There exists $m \in M, H \in {[n] \choose \lfloor q/m \rfloor}, \kappa \in \{0, 1\}^H$ such that, with probability at least $5/6$, $\overline{R}_{m, H, \kappa,b^*}(X) \geq \theta$.
\end{claim}

\begin{proof} [Proof of \pref{lemma:global-decoder} from the soundness and completeness claims]
    By \pref{claim:soundness}, with probability at least $5/6$, no iteration of the algorithm outputs the incorrect bit $1-b^*$.
    By \pref{claim:completeness}, with probability at least $5/6$, there exists an iteration for which the algorithm outputs $b^*$ if it reaches that iteration. 
    
    Therefore, by the union bound, with probability at least $2/3$ both events occur simultaneously.
    On this event, the algorithm cannot output $1-b^*$, and either it halts earlier with output $b^*$, or it reaches the iteration promised by \pref{claim:completeness} and outputs $b^*$.
    
    Hence, with probability at least $2/3$ the algorithm outputs $b^*=x_i$, as required.
\end{proof}

\begin{proof} [Proof of \pref{claim:soundness}, the soundness claim ]
    We first argue that for every fixed $m, H, \kappa$:
    \[
        \Pr[\overline{R}_{m, H, \kappa, (1-b^*)}(X) \geq \theta] \leq \exp \left( -\eta^2\frac{pr}{16q m} \right) \; .
    \]
    
    We claim that $|H| \leq \delta n$.
    By the construction, for some $j \in [d]$ we have $|H| = \left\lfloor \frac{n}{n^{j/d}} \right \rfloor \leq n^{1-1/d}$.
    By \pref{remark:wlog-p} we may assume without loss of generality that $n^{1/d} > Dq^2 \log n$.
    Hence, for sufficiently large $n$ we have 
    \[ 
        |H| \leq n^{1-1/d} \leq \frac{n}{ Dq^2 \log n} \leq \delta n \;, 
    \]
    for any constant $\delta$.

    Notice that $y$ differs from the codeword $c = C(x)$ only on indices that are in $H$.
    Hence, from the bound of the size of $H$, $y$ is $\delta$-close to $c$, and by the soundness guarantee of the relaxed decoder,
    \[
        \Pr_{T \sim \tau}[T(y) = 1-b^*] \leq 1 - \beta \; .
    \]

    Now,
    \[
        \sum_{S \subseteq[n] \setminus H} \rho_{H, \kappa, 1-b^*}(S) = \sum_{S \subseteq[n] \setminus H}\Pr_{T \sim \tau}[T(y) = 1-b^* \ \wedge\ Q_T(y) \setminus H=S] =  \Pr_{T \sim \tau}[T(y) = 1-b^*] \; .
    \]
    That is, $\rho_{H, \kappa, 1-b^*}$ is a sub-distribution with total weight of at most $1-\beta$.
    Hence, the expected value of the (non-capped) $p$-Horvitz--Thompson estimator for  $\rho_{H, \kappa, 1-b^*}$ is at most $1-\beta$.
    By the choice of $\theta$ and $\eta$ we get
    \[
        \E[R_{m, H, \kappa, 1-b^*}(X)] \leq 1-\beta < \theta - \eta/2\; .
    \]

    Now, recall that for every sub-distribution $\mu$ we have $\E[\overline{R}(X)] \leq \E[R(X)]$.
    Hence, by \pref{corollary:cht-concentration}, and plugging $\Gamma = 2m/pr$ and $\gamma = \eta/2$, this implies that:
    \[
        \Pr[\overline{R}_{m, H, \kappa, 1-b^*}(X) \geq \theta]
        \leq \exp \left( -\frac{(\eta/2)^2}{2q \Gamma}\right)
        = \exp  \left( - \eta^2 \frac{pr}{16q m}\right) \; .
    \]

    We finish the proof by applying the union bound three times.
    First, fix $m$ and $H \in {[n] \choose \lfloor q/m \rfloor}$, and iterate over all $\kappa \in \{0, 1\}^H$.
    There are $2^{|H|} = 2^{\left \lfloor q/m \right \rfloor}$ possible assignments $\kappa$.
    Therefore, for every fixed $m$ and $H$, 
    \[
        \Pr[\exists \kappa \in \{0, 1\}^H: \overline{R}_{m, H,\kappa, 1-b^*}(X) \geq \theta] \leq  2^{q/m} \cdot \exp \left( - \eta^2 \frac{pr}{16q m} \right) \; .
    \]
    
    Second, fix $m$, and iterate over all subsets $H \in {[n] \choose \lfloor q/m \rfloor}$.
    There are ${n \choose \lfloor q/m \rfloor} \leq n^{q/m} = 2^{\frac{q}{m}\log n}$ such subsets.
    Hence, for every fixed $m$:
    \[
        \Pr[\exists H \in {[n] \choose \lfloor q/m \rfloor}, \exists \kappa \in \{0, 1\}^H: \overline{R}_{m, H,\kappa, 1-b^*}(X) \geq \theta] \leq  2^{\frac{q}{m}(1+ \log n)} \cdot \exp \left( -\eta^2 \frac{pr}{16q m} \right) \; .
    \]

    Third, we iterate over all $d$ possible values of $m$.
    Using $2^x \leq \exp(x)$, for every $m \in M$ we have
    \[
        2^{\frac{q}{m}(1+\log n)} \cdot \exp\left(- \eta^2 \frac{pr}{16qm}\right) \leq \exp \left(\frac{q(1+\log n)-\eta^2 pr/16q}{m} \right) \; .
    \]
    Assuming $pr > \frac{16q^2(1+\log n)}{\eta^2}$, the numerator in the exponent is negative.
    Since $m \leq q$, we conclude that
    \[
        \Pr[\exists m,H,\kappa: \overline{R}_{m,H,\kappa,1-b^*}(X)\geq \theta] \leq d\exp\left( 1+\log n - \eta^2 \frac{pr}{16q^2} \right) \; .
    \]

    By our choice of parameters,
    \[
        pr = D q^2 \log n \; .
    \]
    Therefore,
    \[
        d\exp\left( 1+\log n - \eta^2 \frac{pr}{16q^2} \right) = d\exp\left( 1+\log n - \frac{\eta^2 D}{16}\log n \right) \; .
    \]

    Since $d=\lfloor (\lambda + \varepsilon)q\rfloor +1\leq (\lambda + \varepsilon)n + 1$, for a sufficiently large constant $D$ the above quantity is smaller than $1/6$.

    Hence, with probability at least $5/6$, for every $m \in M$, every $H \in {[n] \choose \left \lfloor q/m \right \rfloor}$, and every $\kappa \in \{0,1\}^H$,
    \[
        \overline{R}_{m,H,\kappa,1-b^*}(X) < \theta \; ,
    \]
    which proves the claim.
\end{proof}

\begin{proof} [Proof of \pref{claim:completeness}, the completeness claim]

We prove the completeness claim by analyzing the relaxed decoder on the valid codeword $c=C(x)$.

\paragraph{The completeness sub-distribution. }
    Define the sub-distribution over sets
    \[
        \nu(S) = \Pr_{T \sim \tau}[T(c) = b^* \land Q_T(c) = S] \; .
    \]
    In words, $\nu$ can be described by the following process:
    Choose at random a decision tree $T$ according to the distribution $\tau$.
    Then, follow the operation of the tree $T$ on the input $c$.
    If its output bit is not $b^*$, ``ignore'' this tree.
    Otherwise, return the set of indices queried by $T$.

    Now, notice that $\|\nu\|_1$ is exactly the completeness probability of the relaxed decoder:
    \[
        \|\nu\|_1 = \sum_S{\Pr_{T \sim \tau}[T(c) = b^* \land Q_T(c) = S]} = \Pr_{T \sim \tau}[T(c) = b^*] \; .
    \]
    By the completeness guarantee of the relaxed decoder, we therefore have
    \[
        \|\nu\|_1 \geq \alpha \; .
    \]

\paragraph{Extracting spreadness. }
    Each decision tree in $\supp(\tau)$ queries at most $q$ indices, and hence $\nu$ is supported on sets of size at most $q$.
    We apply \pref{corollary:large_spread_daisy2} to $\nu$, with $d = \lfloor (\lambda + \varepsilon)q \rfloor + 1$.
    We deduce there exists $j \in [d]$, $\D^* \subseteq \supp(\nu)$ and $H^* \subseteq [n]$ such that the sub-distribution $(\nu|_{\D^*})^{\circ H^*}$ is $(m^*, n^{1/d})$-spread, where:
    \begin{align*}        
        m^* &= q \frac{n^{j/d}}{n}, &
        |H^*| &= \left\lfloor \frac{n}{n^{j/d}} \right \rfloor = \left\lfloor \frac{q}{m^*} \right \rfloor, &
        \nu(\D^*) &\geq \left( 1 - \frac{q}{d} \right)\|\nu\|_1 \; .
    \end{align*}

\paragraph{The correct guess. }
    Now, define $\kappa^* \in \{0, 1\}^{H^*}$ by $\kappa^*(j)=c_j$.
    With this definition, we have $c_{H^* \xleftarrow{} \kappa^*}=c$.
    Hence, pointwise,
    \[
        \nu^{\circ H^*} =\rho_{H^*, \kappa^*, b^*} \; ,
    \]
    
    and therefore
    \[
        (\nu|_{\D^*})^{\circ H^*} \leq \rho_{H^*, \kappa^*, b^*} \; .
    \]

    We next claim that $m^*, H^*, \kappa^*$ are ``the correct guess'' of the sample-based decoder.
    That is, we argue that the $\frac{2m^*}{pr}$-Capped, $p$-Horvitz--Thompson estimator $\overline{R}_{m^*, H^*, \kappa^*, b^*}$ for the sub-distribution $\rho_{H^*, \kappa^*, b^*}$ satisfies:
    \[
        \Pr[\overline{R}_{m^*, H^*, \kappa^*, b^*}(X) \geq \theta] \geq 5/6 \; .
    \]

\paragraph{From spreadness to a large expected CHT estimate. }
    Denoting $\sigma = (\nu|_{\D^*})^{\circ H^*}$, we have $\| \sigma \|_1 =\nu(\D^*)$.
    Let $R_\sigma$ be the $p$-Horvitz--Thompson estimator for $\sigma$.
    
    Since the uncapped Horvitz--Thompson estimator is unbiased, and by the choice of $\theta$ and $\eta$,
    \begin{align*}
        \E[R_{\sigma}] = \nu(\D^*) 
        \geq \alpha \left(1-\frac{q}{d} \right)
        > \theta + \eta/2  \;.
    \end{align*}
    
    Now, let $\overline{R}_\sigma$ be the $\frac{2m^*}{pr}$-Capped, $p$-Horvitz--Thompson estimator for $\sigma$.
    Since $\sigma$ is $(m^*, n^{1/d})$-spread, and $pr > 2(q-1)$, we get from \pref{lemma:cht-expectation} that
    \[
        \E[R_{\sigma}] - \E[\overline{R}_{\sigma}] = O\left(\frac{q^2}{pr}\right) \; .
    \]
    Since we have $pr = \omega(q^2)$, for sufficiently large $n$ we have $\E[\overline{R}_{\sigma}] \geq \E[R_{\sigma}] -  \frac{\eta}{4} \geq \theta + \frac{\eta}{4}$.

    Next, from the monotonicity of the Capped Horvitz--Thompson estimator, and as $ \rho_{H^*, \kappa^*, b^*} \geq \sigma$, we also have
    \[
        \E[\overline{R}_{m^*, H^*, \kappa^*, b^*}] \geq \E[\overline{R}_{\sigma}] \geq \theta + \frac{\eta}{4} \; .
    \]

    \paragraph{From expectation to high probability. }
    Now, by applying \pref{corollary:cht-concentration} with $\gamma = \frac{\eta}{4}$ and $\Gamma = \frac{2m^*}{pr}$, we get that:
    \begin{align*}        
         \Pr[\overline{R}_{m^*, H^*, \kappa^*, b^*}(X) < \theta] 
         &\leq \exp\left(-\frac{\left(\eta/4 \right )^2}{2q\Gamma} \right) \\
         &= \exp \left( - \frac{\eta^2}{32q \Gamma} \right) \\
         &= \exp \left( - \eta^2 \frac{pr}{64q m^*}  \right) \\
         &\leq \exp \left( - \eta^2 \frac{pr}{64q^2}  \right) \; ,
    \end{align*}
    where the last inequality follows from $m^* \leq q$.

    Now, since $pr = \omega(q^2)$ and $\eta$ is a constant independent of $n$, for sufficiently large $n$ we have 
    \[
        \Pr[\overline{R}_{m^*, H^*, \kappa^*, b^*}(X) < \theta] \leq \frac{1}{6} \; ,
    \]
    which finishes the proof.
    \end{proof}

\subsection{Completing the Proof}
\begin{proof} [Proof of \pref{thm:main}]
    Let $C :\{0,1\}^k\rightarrow \{0, 1\}^n$ be an $(q, \delta, \alpha, \beta)$-RLDC where $\delta, \alpha, \beta$ are constants satisfying $\alpha + \beta > 1$.
    Let $\varepsilon > 0$ be a fixed constant, and denote $\lambda = \frac{\alpha}{\alpha + \beta - 1}$.

    By \pref{lemma:global-decoder}, $C$ has a sample-based decoder with error probability $1/3$ and sampling probability $p= \min\left(1, \frac{Dq^2\log n}{n^{1/d}} \right) \leq \frac{Dq^2\log n}{n^{1/d}}$ for some constant $D$ and $d = \lfloor (\lambda + \varepsilon)q\rfloor + 1$.

    By \pref{lemma:decoder-to-lb}, and absorbing the constant $D$ into the $\Omega$ notation, this implies:
    \[
        \frac{q^2\log n}{n^{1/d}} n = \Omega(k) \; .
    \]
    Now, $d \leq (\lambda + \varepsilon)q + 1$, and so
    \[
        q^2 \log n \cdot n^{1-\frac{1}{(\lambda+\varepsilon)q + 1}} = \Omega(k) \; .
    \]
    We can assume without loss of generality that $n \leq k^2$, since otherwise the lower bound trivially holds.
    Under this assumption, $\log n \leq 2 \log k$.  
    \[
        q^2 \log k \cdot n^{\frac{(\lambda + \varepsilon)q}{(\lambda + \varepsilon)q + 1}} = \Omega(k) \; .
    \]
    By rearranging the terms we get
    \[
    n = \Omega \left(\left(q^{-2} \frac{k}{\log k} \right)^{\frac{(\lambda + \varepsilon)q + 1}{(\lambda + \varepsilon)q}} \right)
    = \Omega \left(\left(q^{-2} \frac{k}{\log k} \right)^{1+ \frac{1}{(\lambda + \varepsilon)q}} \right) \; .
    \]

    Finally, we observe that $\left(q^{-2} \right)^{1/(\lambda+\varepsilon)q} = \Theta(1)$, and hence $\left(q^{-2} \right)^{1+ \frac{1}{(\lambda+\varepsilon)q}} = \Theta(q^{-2})$.
    The theorem follows.
\end{proof}

\section*{Acknowledgments}
We would like to thank Irit Dinur, Oded Goldreich, and Guy Rothblum for valuable discussions and helpful insights about relaxed LDCs.
We also thank Gaia Carenini for a helpful correspondence on spreadness and its applications.

\bibliographystyle{alpha}
\bibliography{references}
\end{document}